\documentclass{aa}  

\usepackage{graphicx}
\usepackage{subfig}
\usepackage{txfonts}
\usepackage[hidelinks, colorlinks=True, citecolor=black, urlcolor=blue, linkcolor=black]{hyperref}
\usepackage{arydshln}
\usepackage{xcolor}
\usepackage{units}
\usepackage{ulem}

\usepackage{natbib}

\begin{document}

    \title{Simulation-guided galaxy evolution inference: A case study with strong lensing galaxies}

   \author{Andreas Filipp
          \inst{1,2,3,4,5} \fnmsep\thanks{\email{andreas.filipp@umontreal.ca}} %(MPA&TUM)?
          \and
          Yiping Shu \inst{6,2}   \fnmsep\thanks{\email{yiping.shu@pmo.ac.cn}}
          \and
          R\"udiger Pakmor \inst{2}
          \and 
          Sherry H. Suyu \inst{1,2,7}
          \and 
          Xiaosheng Huang \inst{8}
          }

   \institute{Technical University Munich, TUM School of Natural Sciences, Physics Department,
                 James-Franck-Strasse 1, 85748 Garching, Germany
            \and
            Max Planck Institute for Astrophysics (MPA), 
                 Karl-Schwarzschlid-Strasse 1, 85748 Garching, Germany
          \and
                Université de Montréal, Physics Department,  
                1375 Ave.Thérèse-Lavoie-Roux, Montréal, QC H2V 0B3
            \and 
                Ciela - Montreal Institute for Astrophysical Data Analysis and Machine Learning, Montréal, Canada
            \and
                Mila - Quebec Artificial Intelligence Institute, Montréal, Canada
            \and
              Purple Mountain Observatory, Chinese Academy of Sciences, Nanjing 210023, People’s Republic of China
            \and
            Academia Sinica Institute of Astronomy and Astrophysics (ASIAA), 11F
of ASMAB, No.1, Section 4, Roosevelt Road, Taipei 10617,
Taiwan
            \and
              University of San Francisco,
              2130 Fulton Street, San Francisco, CA 94117-1080, USA
              }

   \date{Received ...; accepted ...}

  \abstract
   {
   Understanding the evolution of galaxies provides crucial insights into a broad range of aspects in astrophysics, including structure formation and growth, the nature of dark energy and dark matter, baryonic physics, and more. It is, however, infeasible to track the evolutionary processes of individual galaxies in real time given their long timescales. As a result, galaxy evolution analyses have been mostly based on ensembles of galaxies that are supposed to be from the same population according to usually basic and crude observational criteria. We propose a new strategy of evaluating the evolution of an individual galaxy by identifying its descendant galaxies as guided by cosmological simulations. As a proof of concept, we examined the evolution of the total mass distribution of a target strong lensing galaxy at $z=0.884$ using the proposed strategy. We selected 158 galaxies from the IllustrisTNG300 simulation that we identified as analogs of the target galaxy. We followed their descendants and found 11 observed strong lensing galaxies that match in stellar mass and size with the descendants at their redshifts. 
   The observed and simulated results are discussed, although no conclusive assessment is made given the low statistical significance due to the small sample size. Nevertheless, the test confirms that our proposed strategy is already feasible with existing data and simulations. We expect it to play an even more important role in studying galaxy evolution as more strong lens systems and larger simulations become available with the advent of next-generation survey programs and cosmological simulations.}
   
   \keywords{
   Gravitational lensing: strong, Methods: data analysis, Methods: observational, Methods: statistical, Galaxies: evolution, Galaxies: structure 
               }
\maketitle
%
%-------------------------------------------------------------------

\section{Introduction}

Under the current standard cosmological model, galaxies form inside dark-matter halos due to gravitational instability, and grow in mass and size in a bottom-up fashion through mergers \citep[e.g.,][]{White_91_ETG, Kauffmann_93_EGT}. Dominating the massive end of galaxy populations, early-type galaxies (ETGs) are believed to be the end products of this hierarchical merging paradigm. The total mass (i.e., dark matter $+$ baryonic matter) is arguably the single most important property of an ETG, and its distribution across the galaxy contains imprints from various key physical processes that have played a role in the formation and evolution of the galaxy. Dark-matter-only simulations suggest a ``universal'' density profile with a $r^{-1}$ inner profile and a $r^{-3}$ drop-off at large radii, independent of the halo mass \citep[e.g.,][]{Navarro_1997}. Baryonic physics, on the other hand, is found to have substantially reshaped the mass structure of galaxies. In particular, cooling allows gas to condense and form stars in the central regions of galaxies, which leads to the contraction and dark matter \citep[e.g.,][]{Velliscig_2014}. 
Heating due to supernovae or active galactic nucleus feedback, in contrast, can soften the central density concentration \citep[e.g.,][]{Duffy_2010}. In addition, different types of merging events (e.g., dry versus wet and major versus minor) affect the total mass distribution of ETGs in different fashions \citep[e.g.,][]{Ma_2004, Naab_2009, Remus_2013}. The state-of-the-art cosmological simulations predict that the total-density profile slopes (parameterized by the power-law density model of $\rho \propto r^{-\gamma}$) of ETGs will, on average, become shallower from $z \sim 2$ to $z \sim 0.8$ and remain almost constant afterwards \citep[e.g.,][]{Remus_2017_simulated_ETGs, Wang_2019_ETG_profiles_II}. 

Due to the invisible nature of dark matter, observationally measuring the total mass distribution of ETGs turns out to be elusive. So far there have only been two major ways of achieving this, that is through either dynamics or gravitational lensing. In particular, spatially resolved stellar kinematics have been used for measuring the total mass distribution of low- and intermediate-redshift ETGs with $z \leq 0.33$ \citep[e.g.,][]{Lauer_95, Faber_97, SLACS_mass_slope_11, Cappellari_2015}, but extending it to higher redshifts remains technically challenging due to cosmological dimming and apparent size decrease.

Strong gravitational lensing is the other widely used technique for probing the mass structure of ETGs. As a pure gravitational effect, the results of strong lensing are remarkably accurate and measurements of the total mass are robust against different galaxy models and assumptions on galaxy properties. In particular, the total mass enclosed within the Einstein radius can be determined with a few percent uncertainty \citep[e.g.,][]{Koopmans_2006, SLAC_X_Auger_2010, Yiping2016, Sonnenfeld_SL2S_2013}. More importantly, the constraining power of strong lensing does not degrade with galaxy redshift, which makes it a particularly valuable mass probe at high redshifts. \\
In addition, strong lensing is a powerful tool for addressing important cosmological and astrophysical questions, such as dark matter, dark energy, and the expansion of the Universe \citep[e.g.,][]{Treu_2016, Sharma_2022}. For instance, strong gravitational lensing can detect dark matter subhalos \citep[]{Vegetti_2010, Yashar_2016}. Also combining total mass measurements from strong lensing with stellar mass estimations from stellar population synthesis or kinematics indicates that the stellar initial mass function (IMF) is not universal but rather varies with galaxy properties \citep[e.g.,][]{Auger_IMF_2010, Spiniello_IMF_2015}. Making use of the time delays in strong lens systems, independent measurements of the Hubble constant have been obtained \citep[e.g.,][]{Sherry_rectangular_vdisp_2010, Suyu_2013, Millon_2020, Wong_2020, Birrer_2020, Lyne_2022, Shajib2023}, which has been subject to extensive discussions lately \citep[e.g.,][]{Abdalla_2022_Hubblec, Vagnozzi_2022_Hubblec, Riess_2022_hubblec}.

In a strong lens system, the distortion of the background source is tightly related to the lensing potential. Modeling the imaging data of a strong lens system can therefore reconstruct the total mass distribution of the lensing galaxy \citep[e.g.,][]{Tessore2016, ORiordan2020, AndyGu_22}. However, due to known degeneracies, most photometric observables remain unchanged under specific transformations of the total mass distribution, which makes the solution nonunique \citep[e.g.,][]{Falco1985, Schneider2013, Schneider2014}. To break these degeneracies, stellar kinematics of the lensing galaxies have been introduced to further constrain the total mass distribution \citep[e.g.,][]{Grogin_1996, Romanowsky_1999, Blandford_2001, Koopmans_Treu_2002, Koopmans_2003, Warren_2003, Bolton_2006, van_de_Ven_2010, Barnabe_2012, SLACS_mass_slope_11, Akin_2020, Shajib2023}. Such a joint strong lensing and dynamical analysis framework has been successfully applied to over 100 strong lens systems, which are mostly conducted on the Strong Lensing Legacy Survey (SL2S), the Sloan Lens ACS Survey (SLACS), the Baryon Oscillation Spectroscopic Survey (BOSS) Emission-Line Lens Survey (BELLS) -- a subsurvey of BOSS --, and BELLS for the GALaxy-Ly$\alpha$ EmitteR sYstems (BELLS GALLERY) \citep[e.g.,][]{SLAC_X_Auger_2010, Brownstein_2012_use_sel_tec, BELLS_SLACS_mass_slope, Sonnenfeld_SL2S_2013, Yiping2016, Li_2018} up to $z \approx 1$. Most of these analyses to date are based on velocity dispersion measurements in a single aperture or a few bins. Spatially resolved kinematics for the lenses are not yet prevalent, but new facilities such as ground-based adaptive optics assisted integral field units (IFUs) and the James Webb Space Telescope will enlarge the sample of lenses with spatially resolved kinematic data.

Comparing the observed evolution of the total mass distribution of ETGs with cosmological simulations can provide a compelling test on the complex baryonic physics, dark matter-baryon interplay, and galaxy evolution models. However, obtaining accurate evolutionary trends from observations is not straightforward, as it requires identifying an ensemble of observed galaxies that are on the same evolutionary track. To achieve it, the commonly adopted approach so far has been to treat galaxies that satisfy certain observational criteria, which usually do not account for redshift evolution, as from the same population \citep[e.g.,][]{BELLS_SLACS_mass_slope, Sonnenfeld_SL2S_2013}. Such treatments are generally simplistic and can lead to biased conclusions.

In this paper, we propose a new strategy of tracking the evolution of galaxies by making use of cosmological simulations for selecting galaxies on the same evolutionary track, as an attempt of reaching less biased conclusions. Such a simulation-guided galaxy evolution inference also makes the comparison to simulations more straightforward. This paper is structured the following way: We provide a brief summary of the IllustrisTNG project in Section \ref{IllustrisTNG}, which provides the state-of-the-art cosmological magnetohydrodynamic simulations that we choose to use in this work. Details about our proposed simulation-guided galaxy evolution inference are described in Section \ref{ch:method}. As an illustration, the evolution of the total mass distribution of a strong lensing ETG at $z=0.884$ evaluated following the proposed strategy is presented in Section \ref{ch:test_case}. Discussions and conclusions are given in Section \ref{ch:discussions} and Section \ref{ch:Conclusion}.
Throughout this paper we adopt a flat $\Lambda \text{CDM}$ cosmology with $H_0 = \unit[67.74]{km s^{-1} Mpc^{-1}}$, $\Omega_{\rm M} = 0.3089$, split in baryonic and dark matter (with $\Omega_{\rm b} = 0.0486$), and $\Omega_\Lambda = 0.6911$, since these values are used in the IllustrisTNG simulations \citep[][]{IllustrisTNG_cosmology}.

\section{\label{IllustrisTNG}IllustrisTNG simulation}
  
  Big volume simulations aim to reproduce the structure of our universe, based on theoretical assumptions. Comparing the outcomes of simulations to observations allows a direct comparison of theoretical predicted structures and observed structures.

  In the last years simulations moved to bigger volumes and at the same time better spatial and mass resolutions.  
  The IllustrisTNG simulations are currently one of the biggest simulations at highest volumes available. The IllustrisTNG simulations come in three sizes, the TNG50, TNG100, and TNG300. In each case the name indicates the simulated box side-length in comoving Mpc. All simulations were carried out as dark-matter-only, as well as baryonic-matter simulations \citep{Nelson_2019_TNGDR}.

  The results of the IllustrisTNG simulations are found to be in good agreement within limits with observations.
  The simulated structure and galaxy formation, and the overall galaxy morphologies of both elliptical and spiral galaxies agree to a good degree with observations. The galaxy shapes and colors, the stellar populations, the star formation history, the galaxy stellar mass function, the stellar mass to halo mass ratio, and the luminosity of star forming galaxies are found to be close to observations for low and intermediate mass galaxies \citep[e.g.,][]{IllustrisTNG_clustering_2018, Pillepich_2018_TNG_galformation, TNG_sizes_Rodriguez_2019, Lustig_2023}.
  Further, the simulated metallicities and radio emissions are in agreement with observed metallicities on both galaxy and cluster scale within their uncertainties for low and intermediate mass galaxies \citep[e.g.,][]{IllustrisTNG_cosmology, Cannarozzo_2023_metall}. 
  Also on cluster scale the IllustrisTNG simulations shows agreement with observed morthologies \citep[e.g.,][]{Pillepich_2018_cluster}.
  Overall the simulations agree well with observations, but particularly for the most massive galaxies, a deviation from observations can be observed. For example, the simulations lack in number of massive galaxies, and their profile shows discrepancies with observations, see the sources above. Further investigating the discrepancies helps to understand how the simulations can be improved to match better with observations.

  For galaxy evolution studies of massive ETGs, we want to study these most massive galaxies and use as big simulations as possible to get more massive simulated galaxies, but at the same time we want a good mass resolution. We thus chose the IllustrisTNG300 baryonic simulation with a baryonic mass resolution of $1.1\times 10^7 M_\odot$ and a dark matter mass resolution of $5.9\times 10^7 M_\odot$. The softening radius of the TNG300 simulation is $715$pc in comoving coordinates. These mass resolution and softening radius are a good compromise of sufficient resolution and large enough volume for our purposes of studying massive ETGs. The IllustrisTNG simulation data and descriptions are public available on the IllustrisTNG website\footnote{\url{https://www.tng-project.org/}}. 

\section{\label{ch:method}Methodology}
\subsection{Overview}

The ideal way of probing the evolution of an object, is to measure its changes over time. This is, however, infeasible for studying galaxies, because most of the evolutionary processes involved happen on timescales much longer than the human lifetime. As a result, the most common way of quantifying galaxy evolution is by comparing properties of a sample of galaxies at different redshifts. 
These samples are supposed to be from the same population according to some usually basic and crude criteria. While this ``ensemble approximation'' is a feasible alternative, how to properly construct the ensemble is nontrivial, especially as the sample size is growing substantially and the call for less bias becomes increasingly more critical.

We hence propose a new strategy of linking observed galaxies at different redshifts by utilizing evolutionary tracks drawn from state-of-the-art cosmological simulations. More specifically, for any given observed galaxy at redshift $z_0$, we first identify a collection of simulated galaxies at the same redshift that can be considered as its analogs. The simulated descendants (at $z < z_0$) of the analog galaxies are then selected according to the individual merger trees. We use the properties of the simulated descendants to match them to observed galaxies at the same redshift.
The evolution of the target galaxy can be studied by analyzing the so-selected observed galaxies at $z < z_0$. 
Furthermore, by comparing the evolution trends of observed galaxies with simulations, all relevant ingredients adopted in the simulations, such as the cosmological model, dark matter properties, and sub-grid physics, can be tested.

\subsection{Matching between observed and simulated galaxies}
  
  The objects we are studying are massive ETGs, which are found to follow a scaling relation among three basic properties: size, surface brightness, and velocity dispersion, that is the fundamental plane (FP) \citep[e.g.,][]{FP_1998, Koopmans_2006}. Studies have shown that the intrinsic scatter of the FP for massive ETGs is remarkably small ($\lesssim$0.1 (0.07-0.09) dex) up to $z=0.5$ \citep{MIST_FP_00, SDSS_FP_low_z, Joergensen_FP_2013, MASSIVE_ETG_FP}, and also in good agreement with the FP spanned by lensing galaxies \citep{SLAC_X_Auger_2010},
  which suggests that massive ETGs of similar size and surface brightness are likely self-similar in many other aspects.
  We therefore chose to match between observed and simulated galaxies using size and stellar mass, which is a proxy for the surface brightness and more easily available from the simulations. 
 
  To construct the simulated size-stellar mass relation, we selected subhalos, identified by the subfind algorithm \citep{Springel_subfind_01}, with the following properties from the IllustrisTNG300 simulation: 
  \begin{itemize}
      \item total bound stellar mass between $10^{10.9}$ and $10^{12.5}$ $M_\odot$;
      \item star formation rate below 5 $M_\odot$/yr;
      \item less than 40\% of stellar particles with circularity above 0.7.
  \end{itemize}
  The cut on the circularity (the last item) uses a built-in parameter of the simulation to remove clear disk dominated galaxies (described in the data specifications in table C\footnote{\tiny \url{https://www.tng-project.org/data/docs/specifications/}} \citep[][]{illustris_ellept}).
  By default, only the three-dimensional sizes calculated by the subfind algorithm are provided for the subhalos. In order to make a better comparison to observations we projected for every subhalo all its stellar particles along the $z$-axis and compute the projected half-stellar-mass radius. For the stellar mass, we directly used the total stellar mass from the subfind algorithm. 

  To construct the observed size-stellar mass relation, we used galaxies observed by the Sloan Digital Sky Survey (available catalogs were for galaxies from SDSS-I \citep{SDSS_DR7}, and SDSS-III \citep{DR10_SDSS, SDSS_DR12}, but there is no combined catalog for both) 
  for which spectroscopic redshifts are available. The SDSS-I and SDSS-III are complementary in providing a sample of ETGs up to redshift $\approx 0.9$: The SDSS-III BOSS survey focused on higher redshift galaxies with $z>0.45$, which results in a lack of highly massive galaxies at lower redshifts \citep{BOSS_galaxy_redshifts}, but the SDSS-I galaxy sample does not go to high enough redshifts.  
  To obtain a subsample that can be compared to the simulated galaxies, we needed to apply additional cuts on several properties, including stellar mass. For SDSS galaxies, four different estimates for the stellar mass are available, which are computed by four groups using different methods, that are Granada group \citep{DR10_SDSS}, MPA-JHU group\footnote{\url{https://wwwmpa.mpa-garching.mpg.de/SDSS/DR7/}}, Portsmouth group \citep{Portsmouth_Stellar_mass}, and Wisconsin group based on principal component analysis method \citep{Chen_2012} for two different stellar population synthesis models \citep{Wisconsin_synth_03, Wisconsin_syn_2011}. 
  The SL2S and SLACS lens masses use an exponential decay in their star formation history to determine the stellar masses of the galaxies. The only stellar mass catalogs using exponential decay are the MPA-JHU and the Granada group catalogs. To compare the lens galaxies to general ETG galaxies, the star formation history of the catalog should match. Further the flexible stellar population synthesis (FSPS) library used for the Granada catalog is also used for SLACS lenses.
  The MPA-JHU catalog only contains relatively low redshift galaxies, because it only contains galaxies up to DR8. 
  As a consequence of the mentioned points, we used in this work stellar mass from the Granada group. \\
  Similar to the selection of simulated galaxies, we only selected SDSS galaxies that have 
    \begin{itemize}
      \item Granada stellar mass between $10^{10.9}$ and $10^{12.5}$ $M_\odot$;
      \item Granada star formation rate below 5 $M_\odot/yr$;
      \item ellipticity less than 0.3;
      \item error on $i$-band size less than 10\%.
  \end{itemize}
  For galaxy size, we used the effective radius (in the intermediate-axis convention) of the best-fit de-Vaucouleurs model in the $i$ band that is available in the SDSS imaging catalog. 

\begin{table}[htbp]
      \centering
      \small
      \begin{tabular}{c c c c c}
           \hline \hline
           $z$ & observed & y-scatter & simulated & y-scatter \\ \hline
           0.88 & $-4.28+0.44*x$ & 0.14 & $-6.26+0.61*x$ & 0.15 \\
           0.7 & $-5.94+0.58*x$ & 0.17 & $-6.59+0.64*x$ & 0.14 \\
           0.6 & $-5.06+0.50*x$ & 0.15 & $-6.80+0.67*x$ & 0.14 \\
           0.5 & $-5.38+0.53*x$ & 0.14 & $-6.87+0.68*x$ & 0.13 \\
           0.4 & $-6.46+0.62*x$ & 0.14 & $-6.93+0.69*x$ & 0.13 \\
           0.3 & $-6.61+0.64*x$ & 0.13 & $-6.94+0.69*x$ & 0.12 \\
           0.2 & $-5.94+0.59*x$ & 0.14 & $-6.94+0.70*x$ & 0.12 \\
           0.1 & $-6.72+0.66*x$ & 0.14 & $-6.93+0.70*x$ & 0.12 \\ \hline 
      \end{tabular}
      \caption{
      Best-fit size-mass relations
      of observed and simulated galaxies at different redshifts. The variable $x$ is $\text{log}_{10}(M_*/M_\odot)$, 
      and the fitted quantity in columns 2 and 4 are $\text{log}_{10}(R_{\rm e}/{\rm kpc})$. 
      }
      \label{tab:table_slopes}
\end{table}

\begin{figure*}[htbp]
\centering
\includegraphics[width=0.45\textwidth]{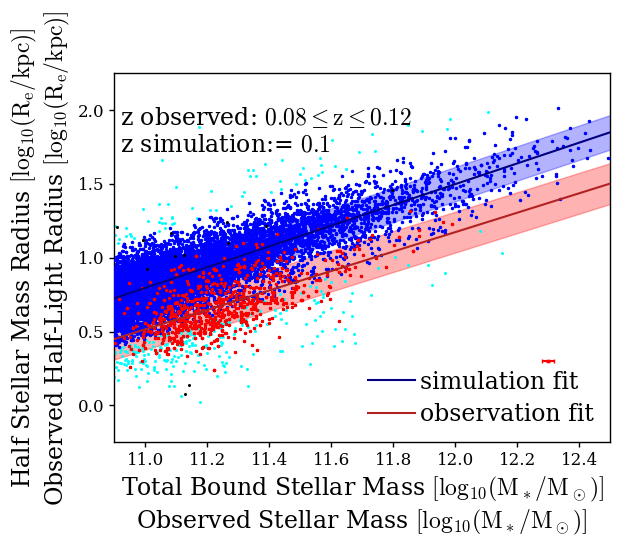}      \includegraphics[width=0.45\textwidth]{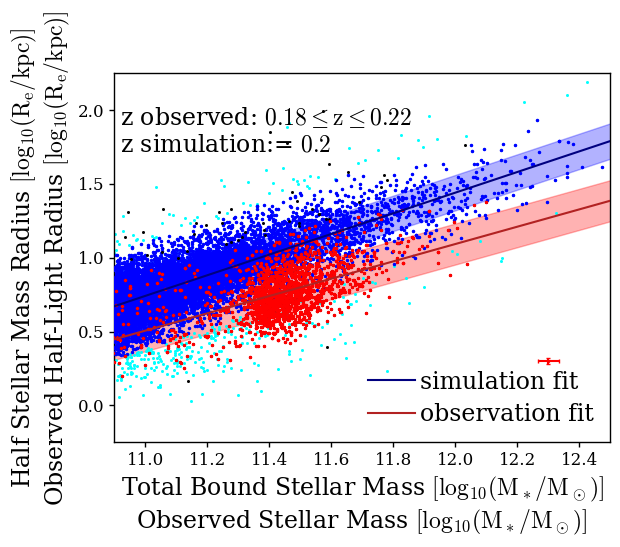}
\includegraphics[width=0.45\textwidth]{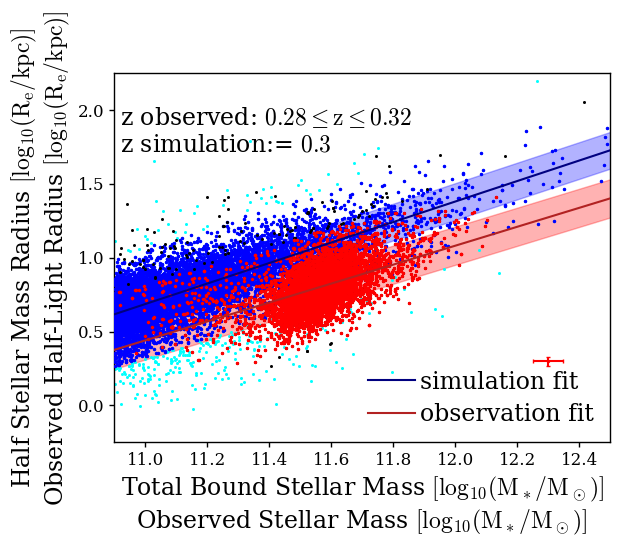}
\includegraphics[width=0.45\textwidth]{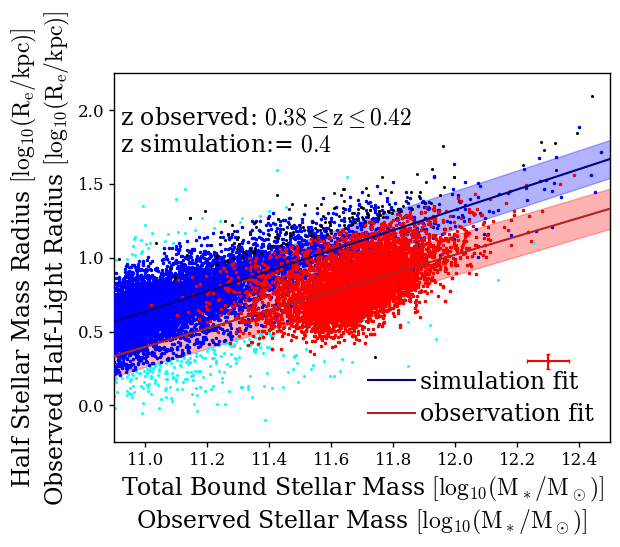}
\includegraphics[width=0.45\textwidth]{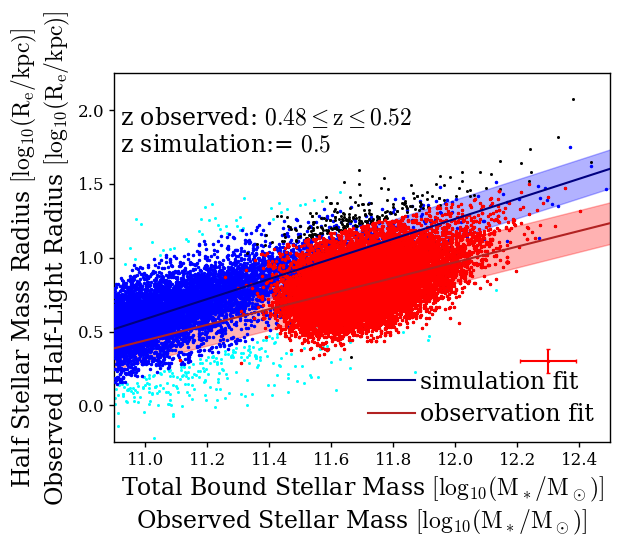}
\includegraphics[width=0.45\textwidth]{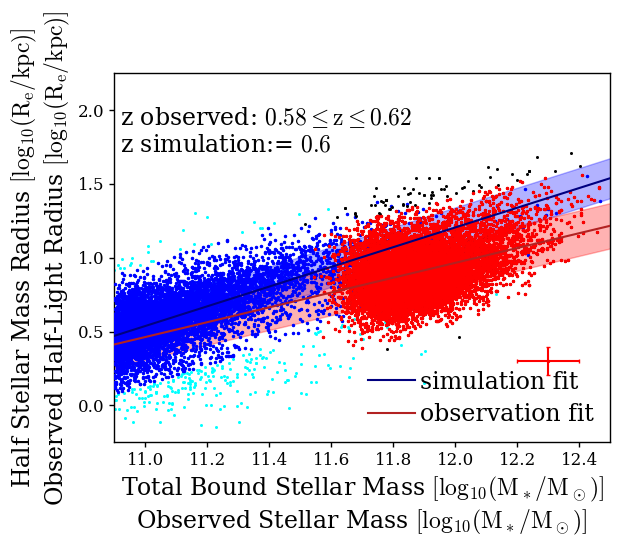}
\caption{Size-stellar mass relations at eight redshift bins from 0.1 to 0.89. The $y$-axis is the half light radius in $i$-band in $\text{log}_{10}(R_{\rm e}/{\rm kpc})$ for observed galaxies or the projected half stellar mass radius in $\text{log}_{10}(R_{\rm e}/{\rm kpc})$ for simulated galaxies. The $x$-axis is the stellar mass in $\text{log}_{10}(M_*/M_\odot)$ for observed galaxies or the total bound stellar mass of a subhalo in $\text{log}_{10}(M_*/M_\odot)$. The observed data points, as well as their best fit, are shown in red, with the mean scatter of the points around the best fit as colored region around it. The median error of the observed data points is shown on the bottom right corner of the plot. The simulated data points and their best fit relation are shown in blue. The cyan colored dots are outliers of the simulated galaxies, identified by $3\sigma$ clipping, that were neglected in the fitting process.}
\label{fig:example_slope}
\end{figure*}
\begin{figure*}[htbp]
\ContinuedFloat
\centering
\includegraphics[width=0.45\textwidth]{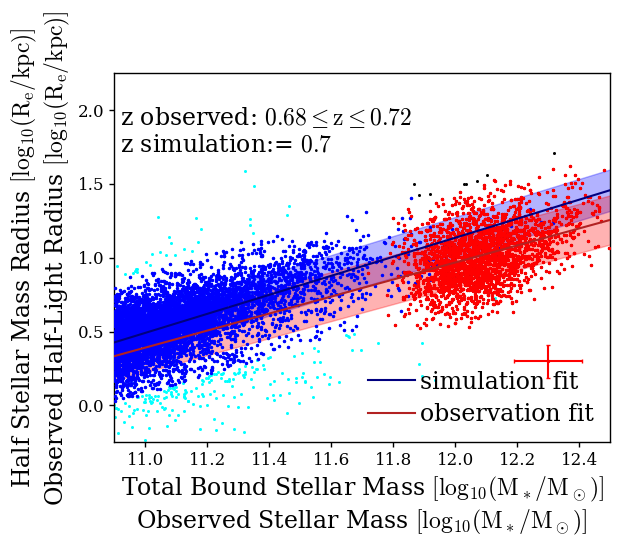}
\includegraphics[width=0.45\textwidth]{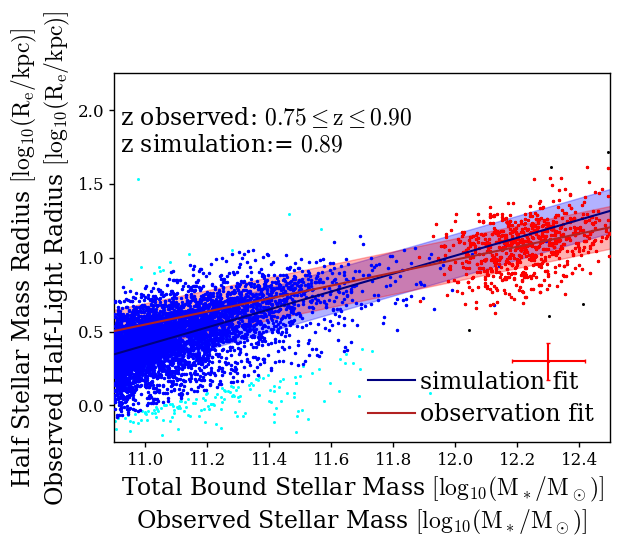}
\caption[]{cont.}
\end{figure*}

Figure~\ref{fig:example_slope} shows the size and stellar mass distributions for simulated and observed galaxies in eight redshift bins from 0.1 to 0.89. The highest redshift bin was chosen to match the redshift of the target galaxy we are analyzing in the next Section. The target observed galaxy has a redshift of $z=0.884$, and the closest redshift in IllustrisTNG is at $z=0.89$. For simulated galaxies, the redshift values are 0.1, 0.2, 0.3, 0.4, 0.5, 0.6, 0.7, and 0.89, while for observed galaxies, the redshift values are $0.1\pm 0.02$, $0.2\pm 0.02$, $0.3\pm 0.02$, $0.4\pm 0.02$, $0.5\pm 0.02$, $0.6\pm 0.02$, $0.7\pm 0.02$, and $0.75 - 0.90$. 
For the highest redshift bin of simulated galaxies at $z=0.89$, we included observed galaxies in the redshift range from 0.75 to 0.90 in order to obtain a large enough sample. 
In each redshift bin, we fitted a linear relation between $\log_{10}(R_{\rm e}/\text{kpc})$ and $\log_{10}(M_*/M_\odot)$ for observed and simulated galaxies separately, and we report the best-fit parameters and the mean scatters in $y$-direction in $\log_{10}(R_{\rm e}/\text{kpc})$ with respect to the best-fit linear relations in Table~\ref{tab:table_slopes}. 

The slopes for the simulated relations are close to 0.7 in all the redshift bins considered, while there is a trend of flattening with cosmic time in the observed size-stellar mass relation. The typical uncertainties on the slope for the simulated data sizes is 0.007 and on the observed fit 0.02. We note that the shaded bands in Figure \ref{fig:example_slope} do not show the uncertainty of the best fit, but the mean scatter of the data points around the best fit. 
We found significant offsets between the observed and simulated size-stellar mass relations, especially in the six lower redshift bins (up to $\approx 0.25$ dex in size). 
This difference has also been observed in previous studies, which suggests it to be primarily caused by offsets in size, because simulated sizes are measured from all star particles while observed sizes are usually measured from profile fitting that may miss the extended, low surface-brightness component, especially in lower redshifts \citep[e.g.,][]{TNG_size_2018_Shy, TNG_sizes_Rodriguez_2019}. 
As a result, in each redshift bin, we chose to use stellar mass and the offset in size (in units of the measured scatter) with respect to the best-fit relation for galaxy matching.

\begin{figure}[t]
\centering
\includegraphics[width=0.45\textwidth]{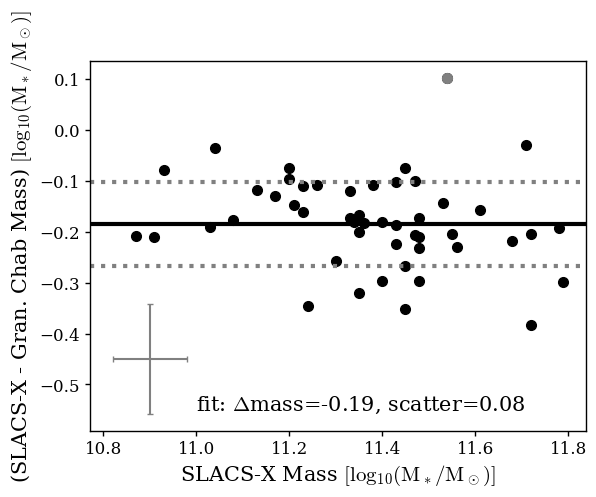}
\caption{Comparison of SLACS X stellar mass with Granada stellar mass with Chabrier IMF. The $y$-axis shows the difference between the SLACS X stellar masses and the Granada masses. The mean difference is found to be $-0.19$, with a scatter of $0.08$. The range of the scatter is indicated by the dotted lines. The errorbar on the left shows the typical uncertainties on the data points. The difference is possibly due to the different assumptions on the stellar initial mass function and the star formation history parameters employed in the models.
}
\label{fig:SLACS_X_Granada}
\end{figure}

\section{\label{ch:test_case}Application on strong lens systems}

As a proof of concept, we constructed an ensemble of strong lens systems at redshifts from $\approx$0.9 to $\approx$0.2 using the proposed strategy and characterized the evolution of their lensing galaxies in terms of the total mass distribution in the central region, particularly in terms of the radial profile of the total mass distribution.
In Section \ref{ch:lens_selection} we describe the selection criteria of the lens systems. Section \ref{ch:mass_slope_calc} gives an overview how we calculated the slope of the radial mass profile of the observed lens systems. In Section \ref{ch:sim_mass_slope} we determine the slope of the radial mass profile of the simulated galaxies, and in Section \ref{ch:results} we show the results of the application of our comparison method on strong lens systems.

\subsection{\label{ch:lens_selection}Lens system selection}

\begin{figure*}[htbp]
\centering
\includegraphics[width=0.45\textwidth]{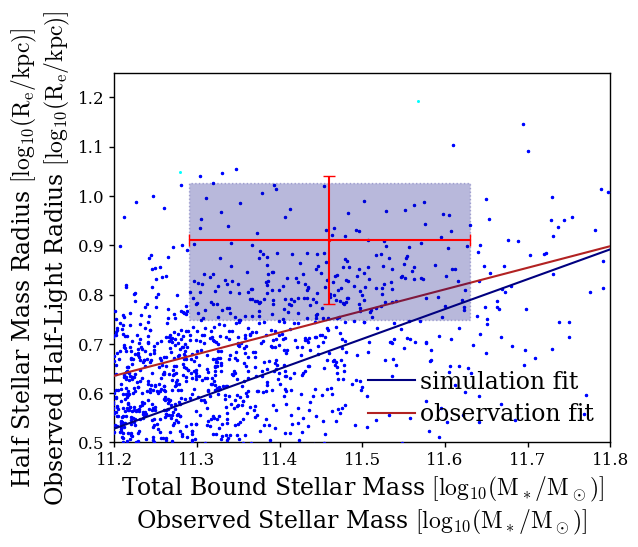}
\includegraphics[width=0.45\textwidth]{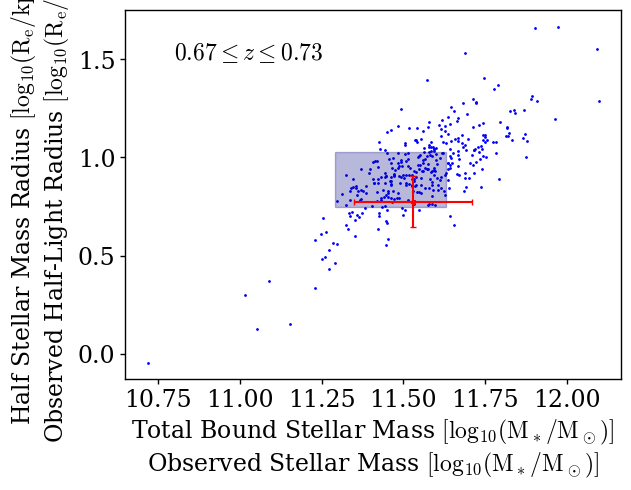}
\caption{Matching simulated and observed galaxies. \textit{Left:} Size-stellar mass distribution of the simulated galaxies at $z=0.89$ (blue) and the target galaxy at redshift $z=0.884$ (red error bar). The 158 simulated galaxies in the blue shaded box are considered as the counterparts of the target galaxy. 
\textit{Right:} Blue points represent the size-stellar mass distribution of the 158 simulated analogs at $z=0.7$. Their size and stellar mass have evolved, as indicated by the blue shaded box that brackets their distributions at $z=0.89$. The red error bar indicates an observed lensing galaxy that was identified as a descendant of the target galaxy at $z=0.7$. All the 158 simulated analogs and the identified observed descendant(s) are used to estimate the {\it simulated} and {\it observed} property (the total mass slope in this work) of the target galaxy at $z=0.7$, respectively.}
\label{fig:selected_galaxies}
\end{figure*}

The strong lens systems were selected from five samples, that are SLACS \citep{SLAC_X_Auger_2010}, BELLS \citep{Brownstein_2012_use_sel_tec}, BELLS GALLERY \citep{Yiping2016}, SLACS for the masses (S4TM) \citep{Yiping_2017}, and SL2S \citep{Sonnenfeld_2013_SL2S_photo, Sonnenfeld_SL2S_2013, Sonnenfeld_SL2S_2015_0884}, that contain in total almost 200 strong lens systems with ETG lensing galaxies at redshifts $\approx 0.05$--$0.9$. 
Since all the lensing galaxies in the first four samples were observed by the SDSS, we used the effective radius in the $i$-band from the SDSS catalog as their sizes, and their stellar masses were directly taken from the Granada catalog. For the SL2S lensing galaxies we used the half-light radii from best-fit de-Vaucouleurs models of the surface brightness of the HST images of the lens galaxies from \citet{Sonnenfeld_2013_SL2S_photo} as their sizes. The stellar masses of the SL2S lensing galaxies were estimated from fitting multiband CFHT photometry using a code implemented by \citet{Auger_2009}, who used this same code to estimate stellar mass for 49 SLACS lensing galaxies. As shown in Figure~\ref{fig:SLACS_X_Granada}, stellar masses estimated by \citet{Auger_2009} are systematically lower than the Granada stellar masses by 0.19 dex on average and the scatter is 0.08 dex\footnote{possibly due to the different assumptions on the stellar initial mass function and the star formation history parameters employed by \citet{Auger_2009} and Granada.}. Because this bias does not vary noticeably with stellar mass, we added 0.19 dex to the reported stellar masses and added in quadrature the 0.08 dex scatter to the reported uncertainties for the SL2S lensing galaxies before galaxy matching.

We selected the lensing galaxy in SL2SJ021801-080247 \citep{Sonnenfeld_SL2S_2015_0884} as our target galaxy based on its highest lens redshift among these five samples, which is at $z = 0.884$. 
The target galaxy has an estimated stellar mass of $\text{{log}}_{10}(M_*/M_\odot)=11.46 \pm 0.17$ (after conversion) and a size of $\text{{log}}_{10}(R_{\text{{e}}}/{\text{{kpc}}})=0.91 \pm 0.13$.
To identify analogs of the target galaxy in the simulations, we applied the following criteria:
\begin{enumerate}
\item The {\it simulated} galaxies have the same stellar mass as the {\it observed} target galaxy within $\pm 1\sigma$ uncertainties, that is logarithmic stellar masses in the range of $11.46 \pm 0.17$;
\item The {\it simulated} galaxies and the {\it observed} target galaxy have the same relative offsets in size (after taking into account the scatters) with respect to the best-fit {\it simulated} and {\it observed} size-stellar mass relation respectively. \\
More specifically,
\begin{enumerate}[i)]
\item At the stellar mass of the target galaxy ($\text{{log}}_{10}(M_*/M_\odot)=11.46$), the best-fit {\it observed} size-stellar mass relation in the $z=0.89$ redshift bin suggests a mean logarithmic size of $0.76$ with a scatter of $0.14$ dex;
\item The {\it observed} $1\sigma$ logarithmic size interval for the target galaxy, that is 0.78--1.04, thus corresponds to 1.13--1.98 times the scatter above the best-fit observed size-stellar mass relation;
\item The best-fit {\it simulated} size-stellar mass relation in the same redshift bin suggests a mean logarithmic size of $0.7306$ with a scatter of $0.15$ dex at the stellar mass of $10^{11.46}M_\odot$ (see Table \ref{tab:table_slopes});
\item We require {\it simulated} galaxies to have logarithmic sizes in the range of 0.7495 ($=0.7306+1.126 \times 0.15$) to 1.0281 ($=0.7306+1.9833 \times 0.15$).
\end{enumerate}

\end{enumerate}
We considered the 158 simulated galaxies that satisfy these criteria as the simulated analogs for the target galaxy, and tracked every individual of them down to z=0 to construct the {\it simulated} evolutionary trends of the target galaxy. The region of the 158 simulated analogs at $z=0.89$ is indicated by the blue shaded box in Figure~\ref{fig:selected_galaxies}.

The same 158 simulated analogs were used to identify {\it observed} descendants of the target galaxy at progressively lower redshift bins. We first followed forward the merger trees of the simulated analogs to a desired redshift bin and updated their sizes and stellar masses. We note that the stellar mass of some simulated galaxies decreased at lower redshifts, likely because they experienced substantial interactions with other objects and lost mass due to tidal forces. We took the simulated galaxies that decrease in mass also into account in our comparison, because the same effect of tidal stripping can happen to observed ETGs. Using the $z=0.7$ bin as an example, the {\it observed} descendants were selected from the lensing galaxies in the five strong lens samples as follows:
\begin{enumerate}
    \item For every {\it simulated} analog, we calculate its relative size offset with respect to the best-fit simulated size-stellar mass relation at $z=0.7$;
    \item For every lensing galaxies within $\Delta z = \pm 0.03$ from the desired redshift bin, we calculate its relative size offset interval with respect to the best-fit observed size-stellar mass relation at $z=0.7$ using its observed $\pm 1\sigma$ logarithmic size interval;
    \item A lensing galaxy is considered as an {\it observed} descendant if its {\it (observed)} relative size offset interval and $\pm 1\sigma$ logarithmic stellar mass interval overlap with at least one simulated analog.
\end{enumerate}
Although all lensing galaxies satisfy the above requirements can be considered as the {\it observed} descendants of the target galaxy, for the purpose of this test case, we further required the selected observed descendants to have similar half-light radius to Einstein radius ratios as SL2SJ021801-080247 (i.e., 1.02) with a tolerance of $\pm 20\%$. This assures that we compare the mass slopes within similar radii of the lensing galaxies. A much larger Einstein radius than half-light radius would measure the profile slope of more dark component than an Einstein radius at the same size of the half-light radius. The tolerance of $20\%$ was found to be a suitable compromise between a large variety in the half-light radius to Einstein radius ratio and a sufficient number of observed analogs. With many more lens systems discovered in the near future, a tighter bound will be possible.
For the considered seven lower redshift bins (see Table~\ref{tab:table_slopes}, 0--4 lensing galaxies satisfy the above requirements, which are used to construct the {\it observed} evolutionary trends of the target galaxy (Table~\ref{tab:lenses_counter}).

    \begin{table*}[h]
    \tiny
        \centering
        \begin{tabular}{l c c c c c c c c c c l} \hline \hline
             lens name & $z_{\rm l}$ & $z_{\rm s}$ & $\theta_{\rm E}['']$ & $R_{\rm eff}['']$ & log$M_*/M_\odot$ & slit$[''\times'']$ & $R_{\rm fib}['']$ & $\sigma_{\rm ap} [\frac{km}{s}]$ &  FWHM$['']$ & $\gamma$ & survey \\ \hline 
             SL2SJ021801-080247 & 0.884 & 2.060 & 1.00 & 1.02 & $11.46 \pm0.15$ & $0.9\times1.60$ &  & $246 \pm 48$ & 1.0 & $1.94\pm 0.25$ & SL2S \\ \hline
             SL2SJ140650+522619 & 0.716 & 1.470 & 0.94 & 0.80 & $11.53\pm 0.17$ & $1.0\times 1.62$ & & $250\pm 20$ & 0.9 & $1.97\pm 0.12$ & SL2S \\ \hdashline
             SL2SJ021411-040502 & 0.609 & 1.880 & 1.41 & 1.21 & $11.53\pm 0.12$ & $1.0\times 1.88$ & & $239\pm 27$ & 0.7 & $1.85\pm 0.07$ & SL2S \\ \hdashline
             SDSSJ0151+0049 & 0.517 & 1.364 & 0.68 & 0.67 & $11.42\pm 0.11$ & & 1.0 & $219\pm 39$ & 1.67 & $2.00\pm 0.14$ & BELLS \\
             SL2SJ142059+563007 & 0.483 & 3.120 & 1.40 & 1.62 & $11.71\pm 0.13$ & $1.0\times 1.62$ & & $228\pm 19$ & 0.8 & $1.95\pm 0.11$ & SL2S \\
             SDSSJ0801+4727 & 0.483 & 1.518 & 0.49 & 0.50 & $11.34\pm 0.09$ & & 1.0 & $98\pm 24$ & 1.67 & $1.64\pm 0.18$ & BELLS \\ \hdashline
             SL2SJ141137+565119 & 0.322 & 1.420 & 0.93 & 0.85 &  $11.23\pm 0.12$ & $1.0\times 1.62$ &  & $214\pm 23$ & 1.3 & $2.16\pm 0.16$ & SL2S \\
             SDSSJ1416+5136 & 0.299 & 0.811 & 1.37 & 1.43 & $11.70\pm 0.05$ & & 1.5 & $240\pm 25$ & 2.18 & $1.90\pm 0.07$ & SLACS \\
             SL2SJ084959-025142 & 0.274 & 2.090 & 1.16 & 1.34 & $11.46\pm 0.12$ & $0.9\times 1.60$ &  & $276\pm 35$ & 0.8 & $2.35\pm 0.15$ & SL2S \\
             SDSSJ1112+0826 & 0.273 & 0.629 & 1.49 & 1.00 & $11.69\pm 0.03$ & & 1.5 & $320\pm 20$ & 3.02 & $1.86\pm 0.05$ & SLACS \\ \hdashline
             SDSSJ1023+4230 & 0.191 & 0.696 & 1.41 & 1.77 & $11.45\pm 0.02$ & & 1.5 & $242\pm 15$ & 1.85 & $2.02\pm 0.05$ & SLACS \\
             SDSSJ1153+4612 & 0.180 & 0.875 & 1.05 & 1.16 & $11.26\pm 0.04$ & & 1.5 & $226\pm 15$ & 1.61 & $2.00 \pm 0.05$ & SLACS \\
              \hline 
        \end{tabular}
        \caption{Lens systems that are selected as possible lower redshift counter parts of system SL2SJ021801-080247. The table shows the lens and source redshifts $z_{\rm l}$ and $z_{\rm s}$, the Einstein radius $\theta_{\rm E}$ in arcseconds, the effective radius $R_{\rm eff}$ in arcseconds, the stellar mass in log$M_*/M_\odot$ (for SL2S lenses adjusted for the comparison), the slit width and length in arcseconds for SL2S systems, the fiber radius $R_{\rm fib}$ in arcseconds for SLACS and BELLS lenses, the measured velocity dispersion within the aperture $\sigma_{\rm ap}$ in $\frac{km}{s}$, the atmospheric seeing FWHM in arcseconds, the derived mass slope $\gamma$ (section \ref{ch:mass_slope_calc}), and the survey name of the lens system. The dashed lines indicate a change of redshift bin. 
        All values, except the mass slope, for the SLACS and BELLS lenses are from \cite{Chen_Yiping_2019}. For the BELLS lenses, the atmospheric seeing is not available for the lens systems used here, so we took the average seeing of those BELLS lenses with values for the seeing. 
        The SL2S slit sizes, velocity dispersions, and seeings are from \cite{Sonnenfeld_SL2S_2013, Sonnenfeld_SL2S_2015_0884}. For systems with more than one velocity dispersion, we took the median of the velocity dispersion values weighted by their signal-to-noise ratio. 
        }
        \label{tab:lenses_counter}
    \end{table*}
  
\subsection{\label{ch:mass_slope_calc}Mass slope calculation of observed galaxies}   
    
    We assumed that the total mass distributions of the lensing galaxies follow a power law profile, that is $\rho (r)= \rho_0 r^{-\gamma}$, which is found to be a good approximation for massive ETGs that were studied in this work \citep[e.g.,][]{Kormann_1994, Suyu_2006, Koopmans_2006, Auger_2009, Sonnenfeld_SL2S_2013, Yiping_2017}. We then inferred $\gamma$ (and $\rho_0$) by comparing the lensing mass constraint from imaging data and the dynamical mass constraint from stellar velocity dispersion measurement. Following \cite{Kormann_1994}, the projected total mass within the Einstein radius (i.e., the lensing mass) is given by
    \begin{equation}
        M_{\rm lensing} = \pi \theta_{\rm E}^2 D_{\rm L}^2 \textstyle\sum_{\rm crit}
    \end{equation}
    where $\theta_{\rm E}$ is the Einstein radius inferred from lens modeling, $D_{\rm L}$ is the angular diameter distance from the observer to the lens and $\sum_{\rm crit}$ is the critical density defined as 
    \begin{equation}
        \textstyle\sum_{\rm crit} = \frac{c^2}{4\pi G} \frac{D_{\rm S}}{D_{\rm L} D_{\rm LS}}
    \end{equation}
    with $c$ the speed of light, $G$ the gravitational constant, $D_{\rm S}$ the angular diameter distance from the observer to the source, and $D_{\rm LS}$ the angular diameter distance between lens and source.

    The observed velocity dispersion is the projected and luminosity-weighted average of the three-dimensional velocity dispersion profile, which is determined by the total mass distribution. Assuming spherical symmetry and following the spherical Jeans equation, one can get the expression for the radial velocity dispersion profile of luminous matter $\sigma ^2(r)$
    \begin{equation}
        \sigma^2(r) = \frac{G \int_r^\infty dr' v(r') M(r') r'^{2\beta -2}}{r^{2\beta} v(r)}
    \end{equation}
    with $G$ the gravitational constant, $r$ the radius, $v(r)$ the luminosity density, $M(r)$ the mass enclosed at radius $r$, and $\beta$ the velocity anisotropy parameter \citep[see][]{Binney_1980}. The equation is valid under the assumption that the ratio of stellar number density to stellar luminosity density is spatially constant. The assumption usually holds within the half-light radius of ETGs. 
    The luminosity density $v(r)$ can be inferred from the observed surface brightness distribution via the Abel integral, and the anisotropy parameter is assumed to be zero.

    The luminosity weighting of the velocity dispersion over the aperture can be expressed as
    \begin{equation}
        \langle \sigma ^2 _{*, ||} \rangle = \frac{ \int_\mathcal{A} I(R) \sigma_\parallel^2(R) * {\rm PSF} \,\, R dR d\theta}{ \int_\mathcal{A} I(R) * {\rm PSF} \,\, R dR d\theta }
        \label{eq:vdisp_mod}
    \end{equation}
    where $I(R)$ is the surface brightness distribution (in our case a de-Vaucouleur profile), $\sigma_\parallel (R)$ is the projected velocity dispersion profile, PSF is the point spread function, and $*$ denotes convolution \citep[see][]{Sherry_rectangular_vdisp_2010}. The velocity dispersions of SL2S lenses were measured from spectra extracted within rectangular slits, so the above integration was done in Cartesian coordinates over the slit width and length. For the remaining lenses, whose velocity dispersions were measured from fiber-fed spectra, a circular aperture was used. The slit width and length for the SL2S lenses, the fiber radius $R_{\rm fib}$ for the other lenses, and the atmospheric seeing (full width half maximum) FWHM are reported in Table \ref{tab:lenses_counter}. We note that the predicted $\sigma_{*, ||}$ is now a function of $\gamma$ alone.

\begin{figure}[t]
        \centering
        \includegraphics[width = 0.45\textwidth]{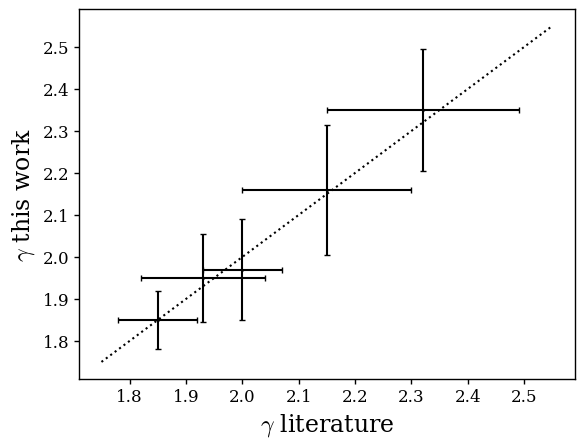}
        \caption{Comparison of the $\gamma$ values of this work with $\gamma$ values from \cite{Sonnenfeld_SL2S_2013}. Our mass slopes agree well with the literature values, so we conclude that our code for determining the mass slope values works well.}
        \label{fig:our_gamma_comp}
    \end{figure}

    The best-fit $\gamma$ value can be inferred by minimizing the $\chi^2$ defined as
    \begin{equation}
        \chi ^2 = \frac{(\sigma_{*,||} (\gamma) - \sigma_{\rm ap} )^2}{\sigma_{\rm unc}^2}
    \end{equation}
    with $\sigma_{*,||} (\gamma)$ as the predicted velocity dispersion for different mass slopes $\gamma$, and $\sigma_{\rm ap}$ as the measured velocity dispersion with uncertainty $\sigma_{\rm unc}$. 
    Since there is only one degree of freedom in the $\chi^2$ function, the 1-$\sigma$ uncertainty on $\gamma$ can be estimated by looking at $\gamma$ values that correspond to $\Delta \chi^2 = 1$ relative to the minimum $\chi^2$ \citep[e.g., see Chapter 7 of the book ``Statistical data analysis'' by ][]{Cowan_1997_chi2}. The derived best-fit $\gamma$ values and their uncertainties are reported in Table~\ref{tab:lenses_counter}. As a consistency check, we show a comparison between our estimated $\gamma$ values and the values estimated by \cite{Sonnenfeld_SL2S_2013} for the five SL2S lenses involved with reported mass slope in figure \ref{fig:our_gamma_comp}, which suggests an almost perfect agreement. The system SL2SJ021801-080247 has no mass slope value provided in literature.

  \begin{figure}[t]
      \centering
      \includegraphics[width=0.45\textwidth]{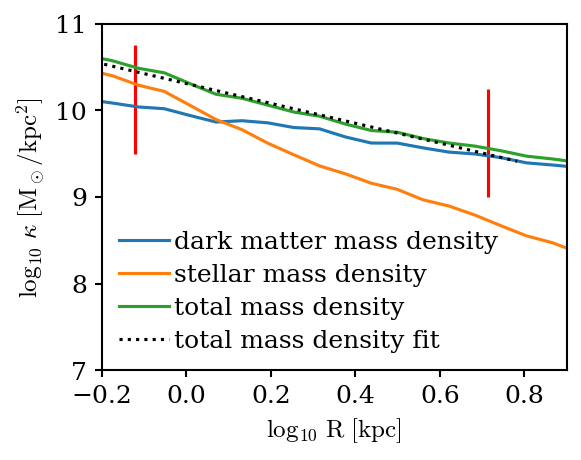}
      \caption{Example for the projected mass density of a simulated galaxy at redshift 0.88 and the power-law fit between the outer and inner radial limit (the red lines). The slope of the fitted line is $-1.146$ and thus the power law mass profile slope $\gamma = 2.146$. 
      }
      \label{fig:simulation_slope}
  \end{figure}

\subsection{\label{ch:sim_mass_slope}Mass slope calculation of simulated galaxies}

  To determine the mass slopes of simulated galaxies, we fitted a power-law profile to their projected mass density distributions. The inner bound of the fit was set to the softening radius of dark matter particles of the Illustris TNG300-1 simulation, which is $1/h\,\text{kpc}= 1.48 \, \text{kpc}$ with $h=0.6774$ in co-moving coordinates \citep[see][]{IllustrisTNG_clustering_2018}. The outer bound of the fit was chosen to be the half stellar-mass radius multiplied by the Einstein-to-effective radius ratio of SL2SJ021801-080247 (i.e., $1.00/1.02$). 
An example for the projected mass density and the power law fit for simulated galaxies is shown in Figure \ref{fig:simulation_slope}. The plot shows the projected dark matter distribution (in blue), the projected stellar mass distribution (in yellow), the projected total mass distribution (in green), and the fit to the total mass distribution between the outer and inner bounds (in black, dotted). The plot is in logarithmic scales for both axes, so the power law fit corresponds to a linear fit.  
  The relation between the mass slope and the total mass distribution is given by
  \begin{equation}
      \bar{\Sigma} (< R) \propto R^{1-\gamma}
  \end{equation}
  with $\gamma$ the mass slope of the profile and $R$ the radius.

  \begin{figure}[t]  
      \centering
      \includegraphics[width=0.45\textwidth]{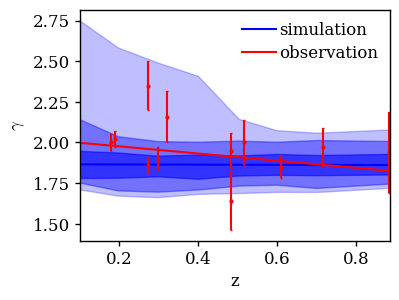}
      \caption{Mass slope comparison of simulated (blue) and observed galaxies (red). The red line is a linear fit to the observed mass slopes over redshift. The blue shaded regions correspond to the 1-$\sigma$, 2-$\sigma$, and 3-$\sigma$ bands inferred from simulated galaxies. 
      }
      \label{fig:gamma_results}
  \end{figure}

\begin{figure}[t]
    \centering
    \includegraphics[width=0.45\textwidth]{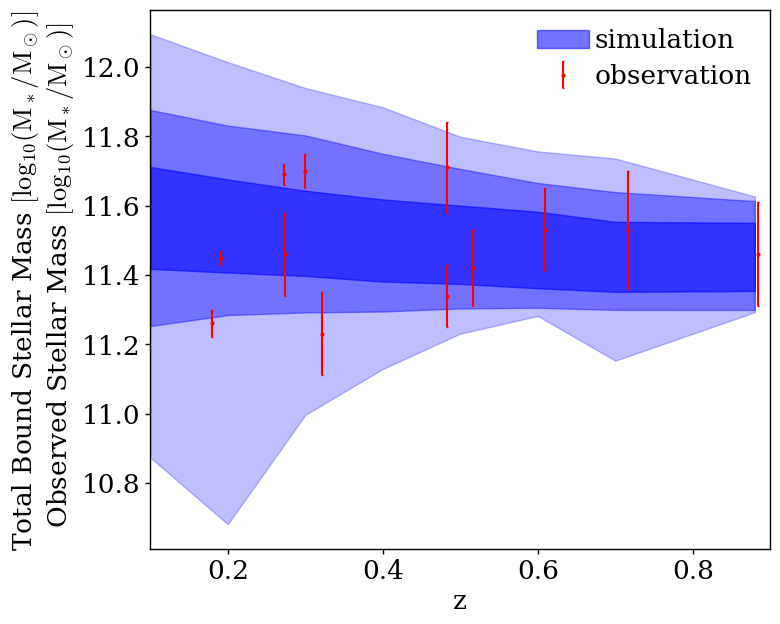}
    \caption{Total bound stellar mass of the simulated galaxies in blue, and the observed stellar mass in red plotted over the redshift. The plot suggests that there is no bias between the two samples.}
    \label{fig:mass_size}
\end{figure}
  
\subsection{\label{ch:results}Test case results}

  Figure \ref{fig:gamma_results} shows the measured total mass-density slopes of the 11 observed lensing galaxies and 158 simulated galaxies that are considered as the descendants of the lensing galaxy in SL2SJ021801-080247. 
  By fitting a line to the measured $\gamma$ values and uncertainties of the 11 observed galaxies, we found that the observed $\gamma$-$z$ relation shows a slightly increasing trend from redshift 0.9 to redshift 0.2 with a slope of $\delta \gamma/ \delta z = -0.22\pm0.18$. For the simulated $\gamma$-$z$ relation, we fitted a line to the mean $\gamma$ values in individual redshift bins using the $1\sigma$ derivation of the mean as uncertainties. We found that the simulated $\gamma$-$z$ relations shows a close-to-constant trend from redshift 0.9 to redshift 0 with a slope of $\delta\gamma/ \delta z = -0.01\pm0.01$. Nevertheless, given the small number of galaxies in each redshift bin, especially on the observation side, these results are by no means conclusive.

  Interestingly, the above evolutionary trends appear to be qualitatively similar with previous findings based on larger samples of galaxies that were usually mass-selected. For example, \citet{BELLS_SLACS_mass_slope} found $\delta\gamma/ \delta z = -0.60 \pm 0.15$ using $\approx 80$ ETG strong lenses at $z \approx 0.1$--$0.7$. \citet{Sonnenfeld_SL2S_2013} found, by taking radius and mass dependence into consideration, the mean evolution for an individual galaxy $d \gamma /dz$ to be on the level of $-0.1$, and \citet{Li_2018} found with similar considerations $\partial \gamma / \partial z$ to be on the level of $-0.3$ using different samples of ETG strong lenses up to $z \approx 1.0$. 
  \citet{Remus_2017_simulated_ETGs} studied galaxies in the Magneticum Pathfinder simulations and found close-to-constant redshift evolution on the total mass-density slope since $z \approx 1$ for ETGs. \citet{Wang_2019_ETG_profiles_II} examined the IllustrisTNG simulations and found similar redshift evolution on the total mass-density slope since $z \approx 1$ for massive ETGs ($> 10^{11} M_\odot$). 

  We further compared the stellar mass evolution of the simulated and observed galaxies in Figure~\ref{fig:mass_size}. According to the 158 simulated analogs, the target galaxy is expected to become slightly more massive over time. From $z=0.88$ to $z=0.1$, the average stellar mass of the simulated analogs increases by 0.1 dex (from a logarithmic stellar mass of 11.45 to 11.55). 
  The stellar mass evolution of the observed descendants agrees with the evolution proposed by the simulation. 
  We chose to not compare the size evolution of observed and simulated galaxies, because the simulated and observed sizes are not directly comparable, as shown in Figure~\ref{fig:example_slope}.

\section{Discussions}
\label{ch:discussions}
  
  The evolution study presented above is clearly limited by the small sample size in both observation and simulation. From the $\approx 200$ strong lens systems used in this work, we only found 0--4 galaxies in each redshift bin (and on average 1.6 galaxies per redshift bin) that are consistent with the descendants of the target galaxy, albeit the relatively loose selection criteria adopted. The situation is not expected to improve much, even if we extend to all currently confirmed strong lens systems with ETGs acting as lenses, which amount to about 400\footnote{According to the Master Lens Database: \\ \url{https://test.masterlens.org/index.php}}. 
  We suggest that an order of magnitude more strong lens systems will be needed in order to obtain a statistically significant result. Thanks to various large imaging surveys and advanced machine learning techniques, a few thousand promising strong lens candidates have recently been discovered \citep[e.g.,][]{Chan_Chitah_15, Xiaosheng_lenses_2020, Lemon_2020, Xiaosheng_lenses_2021, Raoul_2021, Yiping_2022, Tran_2022, Rojas_2022}. Upcoming surveys such as Euclid, Vera C. Rubin Observatory, and Chinese Space Station Telescope will further deliver $\sim 10^5$ new galaxy-scale strong lens systems with high-resolution imaging data in the coming decade \citep[e.g.,][]{Serjeant14, Collett_2015_LSST_DES}. It is equally crucial to carry out follow-up spectroscopic observations that are necessary for confirming the lensing nature and measuring stellar kinematics. 

  Ideally we want to not only compare lenses that are consistent with the descendants, but a set of lenses that have the same distribution in the observables. We have built that in by the requirement of having a similar half-light to Einstein radius ratio. As the number of observed strong lens systems continues growing in the near future, more stringent cuts on these two quantities can be used and additional requirements on properties, such as ellipticity and environment, can also be included to allow an even cleaner selection.
  
  On the simulation side, we found 158 galaxies that can be considered as the analogs of the target galaxy in the IllustrisTNG300 simulation, which is the largest we have access to. Although this appears much better than the situation on the observation side, it is mainly due to the large uncertainties on the stellar mass and size for the target galaxy. If the uncertainties shrink by a factor of 2 (to approximately 0.08 dex in stellar mass and 0.06 dex in size), we could only find 31 analogs from the simulation. This is because known strong lensing galaxies are predominantly massive ETGs ($M_* \gtrsim 10^{11} M_\odot$), which are simply rare. The Illustris team has recently completed the MillenniumTNG simulation\footnote{\url{https://www.mtng-project.org}} \citep[e.g.,][]{MTNG_Barrera, MTNG_Contreras, MTNG_Hadzhiyska, Ruediger_22_Millenium}, which, compared to IllustrisTNG300, has a similar mass resolution but a much larger box with cubic side length of approximately 500 Mpc/h. Given the factor of almost $15$ increase in volume, 
  much better statistics can be obtained when the MillenniumTNG simulation can be used.
  
  Regarding the density slope inference, the uncertainty on $\gamma$ is currently dominated by the velocity dispersion uncertainty, which is typically 10--20\%. More precise inference can be obtained when higher-quality spectroscopic data are available. In addition, more accurate inference is expected when the single velocity dispersion measurement is superseded by two-dimensional stellar kinematics data \citep{SLACS_mass_slope_11}.
  
  Obviously, any observed trend constructed by our proposed strategy is simulation specific, since the observed and simulated galaxy selections depend on the actual merging history, as well as the implemented physics. In other words, the observed trend can be different when a different simulation is used for galaxy matching. As a result, comparing the observed trend constructed this way with the simulated trend can be used to test all relevant ingredients involved in the simulation, such as cosmological models, dark matter properties, and sub-grid physics.

\section{\label{ch:Conclusion}Conclusion}

In this work, we presented a new strategy of probing galaxy evolution by linking observed galaxies at different redshifts according to merger trees drawn from cosmological simulations. The so-linked observed galaxies were considered to be on the same evolutionary path and are therefore used to track how galaxies evolve. Such a simulation-guided galaxy evolution inference is supposed to be less biased than the commonly used approach, that relies on a sample of galaxies that are considered as a single population according to some usually basic and crude observational criteria.

As an illustration, we examined the evolution of the total mass-density slope of massive ETGs acting as strong lenses following this new strategy. We chose the strong lensing galaxy at $z=0.884$ in SL2SJ021801$-$080247 as a starting point and identify 158 simulated galaxies at the same redshift from the IllustrisTNG300 simulation as analogs of the target galaxy according to the stellar mass and size. We then followed the merger trees of the 158 simulated galaxies and successfully found, with the evolution of stellar mass and size taken into account, 11 observed strong lensing galaxies at redshifts of 0.7--0.2 that are analogs of the simulated galaxies and therefore considered descendants of the target galaxy.

The total mass-density slopes of the target galaxy and its 11 descendants were estimated from strong lensing and stellar kinematics data, and the slopes of the 158 simulated galaxies at eight redshift slices from 0.1 to 0.884 were computed directly from the total mass distributions. The observed results suggest the total mass-density slope of the target galaxy will likely increase with time, while the simulated results suggest a mild decreasing trend. Since the current observed results are heavily dominated by the Poisson fluctuation, we choose not to further investigate the apparent different results, which are nonetheless consistent with each other within 1.5$\sigma$.

We would like to emphasize that the proposed simulation-guided galaxy evolution inference is already feasible with currently available strong lens systems and cosmological simulations. In the coming decade, the number of strong lens systems will increase by two orders of magnitude. Cosmological simulations larger than the used IllustrisTNG300 have already been developed and will keep upgrading. The proposed new strategy will take the advantage of this upcoming data explosion and play an even more important role in studying galaxy evolution.

\begin{acknowledgements}
A.F. and S.H.S. thank the Max Planck Society for support through the Max Planck Research Group and the Max Planck Fellowship for S.H.S.. 

A.F. was supported in part from NSERC discovery grant number RGPIN-2020-05102.

Y.S. acknowledges the support from the Max Planck Society and the Alexander von Humboldt Foundation in the framework of the Max Planck-Humboldt Research Award endowed by the Federal Ministry of Education and Research and the China Manned Spaced Project (No. CMS-CSST-2021-A07 and CMS-CSST-2021-A12). 

X.H. acknowledges the University of San Francisco Faculty Development Fund.
\end{acknowledgements}

\bibliographystyle{aa.bst} 
\bibliography{references.bib} %bib

\begin{thebibliography}{100}
\expandafter\ifx\csname natexlab\endcsname\relax\def\natexlab#1{#1}\fi

\bibitem[{{Abazajian} {et~al.}(2009){Abazajian}, {Adelman-McCarthy},
  {Ag{\"u}eros}, {Allam}, {Allende Prieto}, {An}, {Anderson}, {Anderson},
  {Annis}, {Bahcall}, {Bailer-Jones}, {Barentine}, {Bassett}, {Becker},
  {Beers}, {Bell}, {Belokurov}, {Berlind}, {Berman}, {Bernardi}, {Bickerton},
  {Bizyaev}, {Blakeslee}, {Blanton}, {Bochanski}, {Boroski}, {Brewington},
  {Brinchmann}, {Brinkmann}, {Brunner}, {Budav{\'a}ri}, {Carey}, {Carliles},
  {Carr}, {Castander}, {Cinabro}, {Connolly}, {Csabai}, {Cunha}, {Czarapata},
  {Davenport}, {de Haas}, {Dilday}, {Doi}, {Eisenstein}, {Evans}, {Evans},
  {Fan}, {Friedman}, {Frieman}, {Fukugita}, {G{\"a}nsicke}, {Gates},
  {Gillespie}, {Gilmore}, {Gonzalez}, {Gonzalez}, {Grebel}, {Gunn},
  {Gy{\"o}ry}, {Hall}, {Harding}, {Harris}, {Harvanek}, {Hawley}, {Hayes},
  {Heckman}, {Hendry}, {Hennessy}, {Hindsley}, {Hoblitt}, {Hogan}, {Hogg},
  {Holtzman}, {Hyde}, {Ichikawa}, {Ichikawa}, {Im}, {Ivezi{\'c}}, {Jester},
  {Jiang}, {Johnson}, {Jorgensen}, {Juri{\'c}}, {Kent}, {Kessler}, {Kleinman},
  {Knapp}, {Konishi}, {Kron}, {Krzesinski}, {Kuropatkin}, {Lampeitl},
  {Lebedeva}, {Lee}, {Lee}, {French Leger}, {L{\'e}pine}, {Li}, {Lima}, {Lin},
  {Long}, {Loomis}, {Loveday}, {Lupton}, {Magnier}, {Malanushenko},
  {Malanushenko}, {Mandelbaum}, {Margon}, {Marriner}, {Mart{\'\i}nez-Delgado},
  {Matsubara}, {McGehee}, {McKay}, {Meiksin}, {Morrison}, {Mullally}, {Munn},
  {Murphy}, {Nash}, {Nebot}, {Neilsen}, {Newberg}, {Newman}, {Nichol},
  {Nicinski}, {Nieto-Santisteban}, {Nitta}, {Okamura}, {Oravetz}, {Ostriker},
  {Owen}, {Padmanabhan}, {Pan}, {Park}, {Pauls}, {Peoples}, {Percival}, {Pier},
  {Pope}, {Pourbaix}, {Price}, {Purger}, {Quinn}, {Raddick}, {Re Fiorentin},
  {Richards}, {Richmond}, {Riess}, {Rix}, {Rockosi}, {Sako}, {Schlegel},
  {Schneider}, {Scholz}, {Schreiber}, {Schwope}, {Seljak}, {Sesar}, {Sheldon},
  {Shimasaku}, {Sibley}, {Simmons}, {Sivarani}, {Allyn Smith}, {Smith},
  {Smol{\v{c}}i{\'c}}, {Snedden}, {Stebbins}, {Steinmetz}, {Stoughton},
  {Strauss}, {SubbaRao}, {Suto}, {Szalay}, {Szapudi}, {Szkody}, {Tanaka},
  {Tegmark}, {Teodoro}, {Thakar}, {Tremonti}, {Tucker}, {Uomoto}, {Vanden
  Berk}, {Vandenberg}, {Vidrih}, {Vogeley}, {Voges}, {Vogt}, {Wadadekar},
  {Watters}, {Weinberg}, {West}, {White}, {Wilhite}, {Wonders}, {Yanny},
  {Yocum}, {York}, {Zehavi}, {Zibetti}, \& {Zucker}}]{SDSS_DR7}
{Abazajian}, K.~N., {Adelman-McCarthy}, J.~K., {Ag{\"u}eros}, M.~A., {et~al.}
  2009, \apjs, 182, 543

\bibitem[{{Abdalla} {et~al.}(2022){Abdalla}, {Abell{\'a}n}, {Aboubrahim},
  {Agnello}, {Akarsu}, {Akrami}, {Alestas}, {Aloni}, {Amendola}, {Anchordoqui},
  {Anderson}, {Arendse}, {Asgari}, {Ballardini}, {Barger}, {Basilakos},
  {Batista}, {Battistelli}, {Battye}, {Benetti}, {Benisty}, {Berlin}, {de
  Bernardis}, {Berti}, {Bidenko}, {Birrer}, {Blakeslee}, {Boddy}, {Bom},
  {Bonilla}, {Borghi}, {Bouchet}, {Braglia}, {Buchert}, {Buckley-Geer},
  {Calabrese}, {Caldwell}, {Camarena}, {Capozziello}, {Casertano}, {Chen},
  {Chluba}, {Chen}, {Chen}, {Chudaykin}, {Cicoli}, {Copi}, {Courbin},
  {Cyr-Racine}, {Czerny}, {Dainotti}, {D'Amico}, {Davis}, {de Cruz P{\'e}rez},
  {de Haro}, {Delabrouille}, {Denton}, {Dhawan}, {Dienes}, {Di Valentino},
  {Du}, {Eckert}, {Escamilla-Rivera}, {Fert{\'e}}, {Finelli}, {Fosalba},
  {Freedman}, {Frusciante}, {Gazta{\~n}aga}, {Giar{\`e}}, {Giusarma},
  {G{\'o}mez-Valent}, {Handley}, {Harrison}, {Hart}, {Hazra}, {Heavens},
  {Heinesen}, {Hildebrandt}, {Hill}, {Hogg}, {Holz}, {Hooper}, {Hosseininejad},
  {Huterer}, {Ishak}, {Ivanov}, {Jaffe}, {Jang}, {Jedamzik}, {Jimenez},
  {Joseph}, {Joudaki}, {Kamionkowski}, {Karwal}, {Kazantzidis}, {Keeley},
  {Klasen}, {Komatsu}, {Koopmans}, {Kumar}, {Lamagna}, {Lazkoz}, {Lee},
  {Lesgourgues}, {Levi Said}, {Lewis}, {L'Huillier}, {Lucca}, {Maartens},
  {Macri}, {Marfatia}, {Marra}, {Martins}, {Masi}, {Matarrese}, {Mazumdar},
  {Melchiorri}, {Mena}, {Mersini-Houghton}, {Mertens}, {Milakovi{\'c}},
  {Minami}, {Miranda}, {Moreno-Pulido}, {Moresco}, {Mota}, {Mottola}, {Mozzon},
  {Muir}, {Mukherjee}, {Mukherjee}, {Naselsky}, {Nath}, {Nesseris},
  {Niedermann}, {Notari}, {Nunes}, {{\'O} Colg{\'a}in}, {Owens},
  {{\"O}z{\"u}lker}, {Pace}, {Paliathanasis}, {Palmese}, {Pan}, {Paoletti},
  {Perez Bergliaffa}, {Perivolaropoulos}, {Pesce}, {Pettorino}, {Philcox},
  {Pogosian}, {Poulin}, {Poulot}, {Raveri}, {Reid}, {Renzi}, {Riess}, {Sabla},
  {Salucci}, {Salzano}, {Saridakis}, {Sathyaprakash}, {Schmaltz},
  {Sch{\"o}neberg}, {Scolnic}, {Sen}, {Sehgal}, {Shafieloo}, {Sheikh-Jabbari},
  {Silk}, {Silvestri}, {Skara}, {Sloth}, {Soares-Santos}, {Sol{\`a} Peracaula},
  {Songsheng}, {Soriano}, {Staicova}, {Starkman}, {Szapudi}, {Teixeira},
  {Thomas}, {Treu}, {Trott}, {van de Bruck}, {Vazquez}, {Verde}, {Visinelli},
  {Wang}, {Wang}, {Wang}, {Watkins}, {Watson}, {Webb}, {Weiner}, {Weltman},
  {Witte}, {Wojtak}, {Yadav}, {Yang}, {Zhao}, \&
  {Zumalac{\'a}rregui}}]{Abdalla_2022_Hubblec}
{Abdalla}, E., {Abell{\'a}n}, G.~F., {Aboubrahim}, A., {et~al.} 2022, Journal
  of High Energy Astrophysics, 34, 49

\bibitem[{Ahn {et~al.}(2014)Ahn, Alexandroff, \& et~al}]{DR10_SDSS}
Ahn, C.~P., Alexandroff, R., \& et~al, C. A.~P. 2014, The Astrophysical Journal
  Supplement Series, 211, 17

\bibitem[{{Alam} {et~al.}(2015){Alam}, {Albareti}, {Allende Prieto}, {Anders},
  {Anderson}, {Anderton}, {Andrews}, {Armengaud}, {Aubourg}, {Bailey}, {Basu},
  {Bautista}, {Beaton}, {Beers}, {Bender}, {Berlind}, {Beutler}, {Bhardwaj},
  {Bird}, {Bizyaev}, {Blake}, {Blanton}, {Blomqvist}, {Bochanski}, {Bolton},
  {Bovy}, {Shelden Bradley}, {Brandt}, {Brauer}, {Brinkmann}, {Brown},
  {Brownstein}, {Burden}, {Burtin}, {Busca}, {Cai}, {Capozzi}, {Carnero
  Rosell}, {Carr}, {Carrera}, {Chambers}, {Chaplin}, {Chen}, {Chiappini},
  {Chojnowski}, {Chuang}, {Clerc}, {Comparat}, {Covey}, {Croft}, {Cuesta},
  {Cunha}, {da Costa}, {Da Rio}, {Davenport}, {Dawson}, {De Lee}, {Delubac},
  {Deshpande}, {Dhital}, {Dutra-Ferreira}, {Dwelly}, {Ealet}, {Ebelke},
  {Edmondson}, {Eisenstein}, {Ellsworth}, {Elsworth}, {Epstein}, {Eracleous},
  {Escoffier}, {Esposito}, {Evans}, {Fan}, {Fern{\'a}ndez-Alvar}, {Feuillet},
  {Filiz Ak}, {Finley}, {Finoguenov}, {Flaherty}, {Fleming}, {Font-Ribera},
  {Foster}, {Frinchaboy}, {Galbraith-Frew}, {Garc{\'\i}a},
  {Garc{\'\i}a-Hern{\'a}ndez}, {Garc{\'\i}a P{\'e}rez}, {Gaulme}, {Ge},
  {G{\'e}nova-Santos}, {Georgakakis}, {Ghezzi}, {Gillespie}, {Girardi},
  {Goddard}, {Gontcho}, {Gonz{\'a}lez Hern{\'a}ndez}, {Grebel}, {Green},
  {Grieb}, {Grieves}, {Gunn}, {Guo}, {Harding}, {Hasselquist}, {Hawley},
  {Hayden}, {Hearty}, {Hekker}, {Ho}, {Hogg}, {Holley-Bockelmann}, {Holtzman},
  {Honscheid}, {Huber}, {Huehnerhoff}, {Ivans}, {Jiang}, {Johnson},
  {Kinemuchi}, {Kirkby}, {Kitaura}, {Klaene}, {Knapp}, {Kneib}, {Koenig},
  {Lam}, {Lan}, {Lang}, {Laurent}, {Le Goff}, {Leauthaud}, {Lee}, {Lee},
  {Licquia}, {Liu}, {Long}, {L{\'o}pez-Corredoira}, {Lorenzo-Oliveira},
  {Lucatello}, {Lundgren}, {Lupton}, {Mack}, {Mahadevan}, {Maia}, {Majewski},
  {Malanushenko}, {Malanushenko}, {Manchado}, {Manera}, {Mao}, {Maraston},
  {Marchwinski}, {Margala}, {Martell}, {Martig}, {Masters}, {Mathur},
  {McBride}, {McGehee}, {McGreer}, {McMahon}, {M{\'e}nard}, {Menzel},
  {Merloni}, {M{\'e}sz{\'a}ros}, {Miller}, {Miralda-Escud{\'e}}, {Miyatake},
  {Montero-Dorta}, {More}, {Morganson}, {Morice-Atkinson}, {Morrison},
  {Mosser}, {Muna}, {Myers}, {Nandra}, {Newman}, {Neyrinck}, {Nguyen},
  {Nichol}, {Nidever}, {Noterdaeme}, {Nuza}, {O'Connell}, {O'Connell},
  {O'Connell}, {Ogando}, {Olmstead}, {Oravetz}, {Oravetz}, {Osumi}, {Owen},
  {Padgett}, {Padmanabhan}, {Paegert}, {Palanque-Delabrouille}, {Pan},
  {Parejko}, {P{\^a}ris}, {Park}, {Pattarakijwanich}, {Pellejero-Ibanez},
  {Pepper}, {Percival}, {P{\'e}rez-Fournon}, {P{\'e}rez-R{\`a}fols},
  {Petitjean}, {Pieri}, {Pinsonneault}, {Porto de Mello}, {Prada}, {Prakash},
  {Price-Whelan}, {Protopapas}, {Raddick}, {Rahman}, {Reid}, {Rich}, {Rix},
  {Robin}, {Rockosi}, {Rodrigues}, {Rodr{\'\i}guez-Torres}, {Roe}, {Ross},
  {Ross}, {Rossi}, {Ruan}, {Rubi{\~n}o-Mart{\'\i}n}, {Rykoff},
  {Salazar-Albornoz}, {Salvato}, {Samushia}, {S{\'a}nchez}, {Santiago},
  {Sayres}, {Schiavon}, {Schlegel}, {Schmidt}, {Schneider}, {Schultheis},
  {Schwope}, {Sc{\'o}ccola}, {Scott}, {Sellgren}, {Seo}, {Serenelli}, {Shane},
  {Shen}, {Shetrone}, {Shu}, {Silva Aguirre}, {Sivarani}, {Skrutskie},
  {Slosar}, {Smith}, {Sobreira}, {Souto}, {Stassun}, {Steinmetz}, {Stello},
  {Strauss}, {Streblyanska}, {Suzuki}, {Swanson}, {Tan}, {Tayar}, {Terrien},
  {Thakar}, {Thomas}, {Thomas}, {Thompson}, {Tinker}, {Tojeiro}, {Troup},
  {Vargas-Maga{\~n}a}, {Vazquez}, {Verde}, {Viel}, {Vogt}, {Wake}, {Wang},
  {Weaver}, {Weinberg}, {Weiner}, {White}, {Wilson}, {Wisniewski},
  {Wood-Vasey}, {Ye`che}, {York}, {Zakamska}, {Zamora}, {Zasowski}, {Zehavi},
  {Zhao}, {Zheng}, {Zhou}, {Zhou}, {Zou}, \& {Zhu}}]{SDSS_DR12}
{Alam}, S., {Albareti}, F.~D., {Allende Prieto}, C., {et~al.} 2015, \apjs, 219,
  12

\bibitem[{{Auger} {et~al.}(2009){Auger}, {Treu}, {Bolton}, {Gavazzi},
  {Koopmans}, {Marshall}, {Bundy}, \& {Moustakas}}]{Auger_2009}
{Auger}, M.~W., {Treu}, T., {Bolton}, A.~S., {et~al.} 2009, \apj, 705, 1099

\bibitem[{{Auger} {et~al.}(2010{\natexlab{a}}){Auger}, {Treu}, {Bolton},
  {Gavazzi}, {Koopmans}, {Marshall}, {Moustakas}, \&
  {Burles}}]{SLAC_X_Auger_2010}
{Auger}, M.~W., {Treu}, T., {Bolton}, A.~S., {et~al.} 2010{\natexlab{a}}, \apj,
  724, 511

\bibitem[{{Auger} {et~al.}(2010{\natexlab{b}}){Auger}, {Treu}, {Gavazzi},
  {Bolton}, {Koopmans}, \& {Marshall}}]{Auger_IMF_2010}
{Auger}, M.~W., {Treu}, T., {Gavazzi}, R., {et~al.} 2010{\natexlab{b}}, \apjl,
  721, L163

\bibitem[{{Barnab{\`e}} {et~al.}(2011){Barnab{\`e}}, {Czoske}, {Koopmans},
  {Treu}, \& {Bolton}}]{SLACS_mass_slope_11}
{Barnab{\`e}}, M., {Czoske}, O., {Koopmans}, L. V.~E., {Treu}, T., \& {Bolton},
  A.~S. 2011, \mnras, 415, 2215

\bibitem[{{Barnab{\`e}} {et~al.}(2012){Barnab{\`e}}, {Dutton}, {Marshall},
  {Auger}, {Brewer}, {Treu}, {Bolton}, {Koo}, \& {Koopmans}}]{Barnabe_2012}
{Barnab{\`e}}, M., {Dutton}, A.~A., {Marshall}, P.~J., {et~al.} 2012, \mnras,
  423, 1073

\bibitem[{{Barrera} {et~al.}(2022){Barrera}, {Springel}, {White},
  {Hern{\'a}ndez-Aguayo}, {Hernquist}, {Frenk}, {Pakmor}, {Ferlito},
  {Hadzhiyska}, {Delgado}, {Kannan}, \& {Bose}}]{MTNG_Barrera}
{Barrera}, M., {Springel}, V., {White}, S., {et~al.} 2022, arXiv e-prints,
  arXiv:2210.10419

\bibitem[{{Bernardi} {et~al.}(2001){Bernardi}, {Sheth}, {Annis}, {Burles},
  {Eisenstein}, {Finkbeiner}, {Hogg}, {Lupton}, {Schlegel}, {Subbarao},
  {Bahcall}, {Blakeslee}, {Brinkmann}, {Castander}, {Connolly}, {Csabai},
  {Doi}, {Fukugita}, {Frieman}, {Heckman}, {Hennessy}, {Ivezic}, {Knapp},
  {Lamb}, {McKay}, {Munn}, {Nichol}, {Okamura}, {Schneider}, {Thakar}, \&
  {York}}]{SDSS_FP_low_z}
{Bernardi}, M., {Sheth}, R.~K., {Annis}, J., {et~al.} 2001, arXiv e-prints,
  astro

\bibitem[{{Binney}(1980)}]{Binney_1980}
{Binney}, J. 1980, \mnras, 190, 873

\bibitem[{{Birrer} {et~al.}(2020){Birrer}, {Shajib}, {Galan}, {Millon}, {Treu},
  {Agnello}, {Auger}, {Chen}, {Christensen}, {Collett}, {Courbin}, {Fassnacht},
  {Koopmans}, {Marshall}, {Park}, {Rusu}, {Sluse}, {Spiniello}, {Suyu},
  {Wagner-Carena}, {Wong}, {Barnab{\`e}}, {Bolton}, {Czoske}, {Ding},
  {Frieman}, \& {Van de Vyvere}}]{Birrer_2020}
{Birrer}, S., {Shajib}, A.~J., {Galan}, A., {et~al.} 2020, \aap, 643, A165

\bibitem[{{Blandford} {et~al.}(2001){Blandford}, {Surpi}, \&
  {Kundi{\'c}}}]{Blandford_2001}
{Blandford}, R., {Surpi}, G., \& {Kundi{\'c}}, T. 2001, in Astronomical Society
  of the Pacific Conference Series, Vol. 237, Gravitational Lensing: Recent
  Progress and Future Go, ed. T.~G. {Brainerd} \& C.~S. {Kochanek}, 65

\bibitem[{{Bolton} {et~al.}(2012){Bolton}, {Brownstein}, {Kochanek}, {Shu},
  {Schlegel}, {Eisenstein}, {Wake}, {Connolly}, {Maraston}, {Arneson}, \&
  {Weaver}}]{BELLS_SLACS_mass_slope}
{Bolton}, A.~S., {Brownstein}, J.~R., {Kochanek}, C.~S., {et~al.} 2012, \apj,
  757, 82

\bibitem[{{Bolton} {et~al.}(2006){Bolton}, {Burles}, {Koopmans}, {Treu}, \&
  {Moustakas}}]{Bolton_2006}
{Bolton}, A.~S., {Burles}, S., {Koopmans}, L. V.~E., {Treu}, T., \&
  {Moustakas}, L.~A. 2006, \apj, 638, 703

\bibitem[{{Brownstein} {et~al.}(2012){Brownstein}, {Bolton}, {Schlegel},
  {Eisenstein}, {Kochanek}, {Connolly}, {Maraston}, {Pandey}, {Seitz}, {Wake},
  {Wood-Vasey}, {Brinkmann}, {Schneider}, \&
  {Weaver}}]{Brownstein_2012_use_sel_tec}
{Brownstein}, J.~R., {Bolton}, A.~S., {Schlegel}, D.~J., {et~al.} 2012, \apj,
  744, 41

\bibitem[{{Bruzual} \& {Charlot}(2003)}]{Wisconsin_synth_03}
{Bruzual}, G. \& {Charlot}, S. 2003, \mnras, 344, 1000

\bibitem[{{Ca{\~n}ameras} {et~al.}(2021){Ca{\~n}ameras}, {Schuldt}, {Shu},
  {Suyu}, {Taubenberger}, {Meinhardt}, {Leal-Taix{\'e}}, {Chao}, {Inoue},
  {Jaelani}, \& {More}}]{Raoul_2021}
{Ca{\~n}ameras}, R., {Schuldt}, S., {Shu}, Y., {et~al.} 2021, \aap, 653, L6

\bibitem[{{Cannarozzo} {et~al.}(2023){Cannarozzo}, {Leauthaud}, {Oyarz{\'u}n},
  {Nipoti}, {Diemer}, {Huang}, {Rodriguez-Gomez}, {Sonnenfeld}, \&
  {Bundy}}]{Cannarozzo_2023_metall}
{Cannarozzo}, C., {Leauthaud}, A., {Oyarz{\'u}n}, G.~A., {et~al.} 2023, \mnras,
  520, 5651

\bibitem[{{Cappellari} {et~al.}(2015){Cappellari}, {Romanowsky}, {Brodie},
  {Forbes}, {Strader}, {Foster}, {Kartha}, {Pastorello}, {Pota}, {Spitler},
  {Usher}, \& {Arnold}}]{Cappellari_2015}
{Cappellari}, M., {Romanowsky}, A.~J., {Brodie}, J.~P., {et~al.} 2015, \apjl,
  804, L21

\bibitem[{{Chan} {et~al.}(2015){Chan}, {Suyu}, {Chiueh}, {More}, {Marshall},
  {Coupon}, {Oguri}, \& {Price}}]{Chan_Chitah_15}
{Chan}, J. H.~H., {Suyu}, S.~H., {Chiueh}, T., {et~al.} 2015, \apj, 807, 138

\bibitem[{{Chen} {et~al.}(2019){Chen}, {Li}, {Shu}, \&
  {Cao}}]{Chen_Yiping_2019}
{Chen}, Y., {Li}, R., {Shu}, Y., \& {Cao}, X. 2019, \mnras, 488, 3745

\bibitem[{{Chen} {et~al.}(2012){Chen}, {Kauffmann}, {Tremonti}, {White},
  {Heckman}, {Kova{\v{c}}}, {Bundy}, {Chisholm}, {Maraston}, {Schneider},
  {Bolton}, {Weaver}, \& {Brinkmann}}]{Chen_2012}
{Chen}, Y.-M., {Kauffmann}, G., {Tremonti}, C.~A., {et~al.} 2012, \mnras, 421,
  314

\bibitem[{{Collett}(2015)}]{Collett_2015_LSST_DES}
{Collett}, T.~E. 2015, \apj, 811, 20

\bibitem[{{Contreras} {et~al.}(2023){Contreras}, {Angulo}, {Springel}, {White},
  {Hadzhiyska}, {Hernquist}, {Pakmor}, {Kannan}, {Hern{\'a}ndez-Aguayo},
  {Barrera}, {Ferlito}, {Delgado}, {Bose}, \& {Frenk}}]{MTNG_Contreras}
{Contreras}, S., {Angulo}, R.~E., {Springel}, V., {et~al.} 2023, \mnras, 524,
  2489

\bibitem[{{Cowan}(1997)}]{Cowan_1997_chi2}
{Cowan}, G. 1997, {Statistical data analysis}

\bibitem[{{Duffy} {et~al.}(2010){Duffy}, {Schaye}, {Kay}, {Dalla Vecchia},
  {Battye}, \& {Booth}}]{Duffy_2010}
{Duffy}, A.~R., {Schaye}, J., {Kay}, S.~T., {et~al.} 2010, \mnras, 405, 2161

\bibitem[{{Faber} {et~al.}(1997){Faber}, {Tremaine}, {Ajhar}, {Byun},
  {Dressler}, {Gebhardt}, {Grillmair}, {Kormendy}, {Lauer}, \&
  {Richstone}}]{Faber_97}
{Faber}, S.~M., {Tremaine}, S., {Ajhar}, E.~A., {et~al.} 1997, \aj, 114, 1771

\bibitem[{{Falco} {et~al.}(1985){Falco}, {Gorenstein}, \&
  {Shapiro}}]{Falco1985}
{Falco}, E.~E., {Gorenstein}, M.~V., \& {Shapiro}, I.~I. 1985, \apjl, 289, L1

\bibitem[{{Genel} {et~al.}(2015){Genel}, {Fall}, {Hernquist}, {Vogelsberger},
  {Snyder}, {Rodriguez-Gomez}, {Sijacki}, \& {Springel}}]{illustris_ellept}
{Genel}, S., {Fall}, S.~M., {Hernquist}, L., {et~al.} 2015, \apjl, 804, L40

\bibitem[{{Genel} {et~al.}(2018){Genel}, {Nelson}, {Pillepich}, {Springel},
  {Pakmor}, {Weinberger}, {Hernquist}, {Naiman}, {Vogelsberger}, {Marinacci},
  \& {Torrey}}]{TNG_size_2018_Shy}
{Genel}, S., {Nelson}, D., {Pillepich}, A., {et~al.} 2018, \mnras, 474, 3976

\bibitem[{{Grogin} \& {Narayan}(1996)}]{Grogin_1996}
{Grogin}, N.~A. \& {Narayan}, R. 1996, \apj, 464, 92

\bibitem[{{Gu} {et~al.}(2022){Gu}, {Huang}, {Sheu}, {Aldering}, {Bolton},
  {Boone}, {Dey}, {Filipp}, {Jullo}, {Perlmutter}, {Rubin}, {Schlafly},
  {Schlegel}, {Shu}, \& {Suyu}}]{AndyGu_22}
{Gu}, A., {Huang}, X., {Sheu}, W., {et~al.} 2022, \apj, 935, 49

\bibitem[{{Hadzhiyska} {et~al.}(2023){Hadzhiyska}, {Eisenstein}, {Hernquist},
  {Pakmor}, {Bose}, {Delgado}, {Contreras}, {Kannan}, {White}, {Springel},
  {Frenk}, {Hern{\'a}ndez-Aguayo}, {Barrera}, \& {Monica}}]{MTNG_Hadzhiyska}
{Hadzhiyska}, B., {Eisenstein}, D., {Hernquist}, L., {et~al.} 2023, \mnras,
  524, 2507

\bibitem[{{Hezaveh} {et~al.}(2016){Hezaveh}, {Dalal}, {Marrone}, {Mao},
  {Morningstar}, {Wen}, {Blandford}, {Carlstrom}, {Fassnacht}, {Holder},
  {Kemball}, {Marshall}, {Murray}, {Perreault Levasseur}, {Vieira}, \&
  {Wechsler}}]{Yashar_2016}
{Hezaveh}, Y.~D., {Dalal}, N., {Marrone}, D.~P., {et~al.} 2016, \apj, 823, 37

\bibitem[{{Huang} {et~al.}(2021){Huang}, {Storfer}, {Gu}, {Ravi}, {Pilon},
  {Sheu}, {Venguswamy}, {Banka}, {Dey}, {Landriau}, {Lang}, {Meisner},
  {Moustakas}, {Myers}, {Sajith}, {Schlafly}, \&
  {Schlegel}}]{Xiaosheng_lenses_2021}
{Huang}, X., {Storfer}, C., {Gu}, A., {et~al.} 2021, \apj, 909, 27

\bibitem[{{Huang} {et~al.}(2020){Huang}, {Storfer}, {Ravi}, {Pilon}, {Domingo},
  {Schlegel}, {Bailey}, {Dey}, {Gupta}, {Herrera}, {Juneau}, {Landriau},
  {Lang}, {Meisner}, {Moustakas}, {Myers}, {Schlafly}, {Valdes}, {Weaver},
  {Yang}, \& {Y{\`e}che}}]{Xiaosheng_lenses_2020}
{Huang}, X., {Storfer}, C., {Ravi}, V., {et~al.} 2020, \apj, 894, 78

\bibitem[{J{\o}rgensen \& Chiboucas(2013)}]{Joergensen_FP_2013}
J{\o}rgensen, I. \& Chiboucas, K. 2013, The Astronomical Journal, 145, 77

\bibitem[{{Kauffmann} {et~al.}(1993){Kauffmann}, {White}, \&
  {Guiderdoni}}]{Kauffmann_93_EGT}
{Kauffmann}, G., {White}, S.~D.~M., \& {Guiderdoni}, B. 1993, \mnras, 264, 201

\bibitem[{{Koopmans} \& {Treu}(2002)}]{Koopmans_Treu_2002}
{Koopmans}, L. V.~E. \& {Treu}, T. 2002, \apjl, 568, L5

\bibitem[{{Koopmans} \& {Treu}(2003)}]{Koopmans_2003}
{Koopmans}, L. V.~E. \& {Treu}, T. 2003, \apj, 583, 606

\bibitem[{{Koopmans} {et~al.}(2006){Koopmans}, {Treu}, {Bolton}, {Burles}, \&
  {Moustakas}}]{Koopmans_2006}
{Koopmans}, L. V.~E., {Treu}, T., {Bolton}, A.~S., {Burles}, S., \&
  {Moustakas}, L.~A. 2006, \apj, 649, 599

\bibitem[{{Kormann} {et~al.}(1994){Kormann}, {Schneider}, \&
  {Bartelmann}}]{Kormann_1994}
{Kormann}, R., {Schneider}, P., \& {Bartelmann}, M. 1994, \aap, 284, 285

\bibitem[{{La Barbera} {et~al.}(2000){La Barbera}, {Busarello}, \&
  {Capaccioli}}]{MIST_FP_00}
{La Barbera}, F., {Busarello}, G., \& {Capaccioli}, M. 2000, \aap, 362, 851

\bibitem[{{Lauer} {et~al.}(1995){Lauer}, {Ajhar}, {Byun}, {Dressler}, {Faber},
  {Grillmair}, {Kormendy}, {Richstone}, \& {Tremaine}}]{Lauer_95}
{Lauer}, T.~R., {Ajhar}, E.~A., {Byun}, Y.~I., {et~al.} 1995, \aj, 110, 2622

\bibitem[{{Lemon} {et~al.}(2020){Lemon}, {Auger}, {McMahon}, {Anguita},
  {Apostolovski}, {Chen}, {Fassnacht}, {Melo}, {Motta}, {Shajib}, {Treu},
  {Agnello}, {Buckley-Geer}, {Schechter}, {Birrer}, {Collett}, {Courbin},
  {Rusu}, {Abbott}, {Allam}, {Annis}, {Avila}, {Bertin}, {Brooks}, {Burke},
  {Carnero Rosell}, {Carrasco Kind}, {Carretero}, {Costanzi}, {da Costa}, {De
  Vicente}, {Desai}, {Eifler}, {Flaugher}, {Frieman}, {Garc{\'\i}a-Bellido},
  {Gaztanaga}, {Gerdes}, {Gruen}, {Gruendl}, {Gschwend}, {Gutierrez},
  {Honscheid}, {James}, {Kim}, {Krause}, {Kuehn}, {Kuropatkin}, {Lahav},
  {Lima}, {Lin}, {Maia}, {March}, {Marshall}, {Menanteau}, {Miquel}, {Palmese},
  {Paz-Chinch{\'o}n}, {Plazas}, {Roodman}, {Sanchez}, {Schubnell}, {Serrano},
  {Smith}, {Soares-Santos}, {Suchyta}, {Tarle}, \& {Walker}}]{Lemon_2020}
{Lemon}, C., {Auger}, M.~W., {McMahon}, R., {et~al.} 2020, \mnras, 494, 3491

\bibitem[{{Li} {et~al.}(2018){Li}, {Shu}, \& {Wang}}]{Li_2018}
{Li}, R., {Shu}, Y., \& {Wang}, J. 2018, \mnras, 480, 431

\bibitem[{Loureiro {et~al.}(2019)Loureiro, Moraes, Abdalla, Cuceu, McLeod,
  Whiteway, Balan, Benoit-Lévy, Lahav, Manera, Rollins, \&
  Xavier}]{BOSS_galaxy_redshifts}
Loureiro, A., Moraes, B., Abdalla, F., {et~al.} 2019, Monthly Notices of the
  Royal Astronomical Society, 485

\bibitem[{{Lustig} {et~al.}(2023){Lustig}, {Strazzullo}, {Remus}, {D'Eugenio},
  {Daddi}, {Burkert}, {De Lucia}, {Delvecchio}, {Dolag}, {Fontanot}, {Gobat},
  {Mohr}, {Onodera}, {Pannella}, \& {Pillepich}}]{Lustig_2023}
{Lustig}, P., {Strazzullo}, V., {Remus}, R.-S., {et~al.} 2023, \mnras, 518,
  5953

\bibitem[{{Ma} \& {Boylan-Kolchin}(2004)}]{Ma_2004}
{Ma}, C.-P. \& {Boylan-Kolchin}, M. 2004, \prl, 93, 021301

\bibitem[{{Magorrian} {et~al.}(1998){Magorrian}, {Tremaine}, {Richstone},
  {Bender}, {Bower}, {Dressler}, {Faber}, {Gebhardt}, {Green}, {Grillmair},
  {Kormendy}, \& {Lauer}}]{FP_1998}
{Magorrian}, J., {Tremaine}, S., {Richstone}, D., {et~al.} 1998, \aj, 115, 2285

\bibitem[{{Maraston} \& {Str{\"o}mb{\"a}ck}(2011)}]{Wisconsin_syn_2011}
{Maraston}, C. \& {Str{\"o}mb{\"a}ck}, G. 2011, \mnras, 418, 2785

\bibitem[{{Marinacci} {et~al.}(2018){Marinacci}, {Vogelsberger}, {Pakmor},
  {Torrey}, {Springel}, {Hernquist}, {Nelson}, {Weinberger}, {Pillepich},
  {Naiman}, \& {Genel}}]{IllustrisTNG_cosmology}
{Marinacci}, F., {Vogelsberger}, M., {Pakmor}, R., {et~al.} 2018, \mnras, 480,
  5113

\bibitem[{{Millon} {et~al.}(2020){Millon}, {Galan}, {Courbin}, {Treu}, {Suyu},
  {Ding}, {Birrer}, {Chen}, {Shajib}, {Sluse}, {Wong}, {Agnello}, {Auger},
  {Buckley-Geer}, {Chan}, {Collett}, {Fassnacht}, {Hilbert}, {Koopmans},
  {Motta}, {Mukherjee}, {Rusu}, {Sonnenfeld}, {Spiniello}, \& {Van de
  Vyvere}}]{Millon_2020}
{Millon}, M., {Galan}, A., {Courbin}, F., {et~al.} 2020, \aap, 639, A101

\bibitem[{{Naab} {et~al.}(2009){Naab}, {Johansson}, \& {Ostriker}}]{Naab_2009}
{Naab}, T., {Johansson}, P.~H., \& {Ostriker}, J.~P. 2009, \apjl, 699, L178

\bibitem[{{Navarro} {et~al.}(1997){Navarro}, {Frenk}, \&
  {White}}]{Navarro_1997}
{Navarro}, J.~F., {Frenk}, C.~S., \& {White}, S. D.~M. 1997, \apj, 490, 493

\bibitem[{{Nelson} {et~al.}(2019){Nelson}, {Springel}, {Pillepich},
  {Rodriguez-Gomez}, {Torrey}, {Genel}, {Vogelsberger}, {Pakmor}, {Marinacci},
  {Weinberger}, {Kelley}, {Lovell}, {Diemer}, \&
  {Hernquist}}]{Nelson_2019_TNGDR}
{Nelson}, D., {Springel}, V., {Pillepich}, A., {et~al.} 2019, Computational
  Astrophysics and Cosmology, 6, 2

\bibitem[{{O'Riordan} {et~al.}(2020){O'Riordan}, {Warren}, \&
  {Mortlock}}]{ORiordan2020}
{O'Riordan}, C.~M., {Warren}, S.~J., \& {Mortlock}, D.~J. 2020, \mnras, 496,
  3424

\bibitem[{{Pakmor} {et~al.}(2023){Pakmor}, {Springel}, {Coles}, {Guillet},
  {Pfrommer}, {Bose}, {Barrera}, {Delgado}, {Ferlito}, {Frenk}, {Hadzhiyska},
  {Hern{\'a}ndez-Aguayo}, {Hernquist}, {Kannan}, \&
  {White}}]{Ruediger_22_Millenium}
{Pakmor}, R., {Springel}, V., {Coles}, J.~P., {et~al.} 2023, \mnras, 524, 2539

\bibitem[{{Pillepich} {et~al.}(2018{\natexlab{a}}){Pillepich}, {Nelson},
  {Hernquist}, {Springel}, {Pakmor}, {Torrey}, {Weinberger}, {Genel}, {Naiman},
  {Marinacci}, \& {Vogelsberger}}]{Pillepich_2018_cluster}
{Pillepich}, A., {Nelson}, D., {Hernquist}, L., {et~al.} 2018{\natexlab{a}},
  \mnras, 475, 648

\bibitem[{{Pillepich} {et~al.}(2018{\natexlab{b}}){Pillepich}, {Springel},
  {Nelson}, {Genel}, {Naiman}, {Pakmor}, {Hernquist}, {Torrey}, {Vogelsberger},
  {Weinberger}, \& {Marinacci}}]{Pillepich_2018_TNG_galformation}
{Pillepich}, A., {Springel}, V., {Nelson}, D., {et~al.} 2018{\natexlab{b}},
  \mnras, 473, 4077

\bibitem[{{Quenneville} {et~al.}(2022){Quenneville}, {Blakeslee}, {Ma},
  {Greene}, {Gwyn}, {Ciccone}, \& {Nyiri}}]{MASSIVE_ETG_FP}
{Quenneville}, M.~E., {Blakeslee}, J.~P., {Ma}, C.-P., {et~al.} 2022, arXiv
  e-prints, arXiv:2210.08043

\bibitem[{{Remus} {et~al.}(2013){Remus}, {Burkert}, {Dolag}, {Johansson},
  {Naab}, {Oser}, \& {Thomas}}]{Remus_2013}
{Remus}, R.-S., {Burkert}, A., {Dolag}, K., {et~al.} 2013, \apj, 766, 71

\bibitem[{{Remus} {et~al.}(2017){Remus}, {Dolag}, {Naab}, {Burkert},
  {Hirschmann}, {Hoffmann}, \& {Johansson}}]{Remus_2017_simulated_ETGs}
{Remus}, R.-S., {Dolag}, K., {Naab}, T., {et~al.} 2017, \mnras, 464, 3742

\bibitem[{{Riess} {et~al.}(2022){Riess}, {Yuan}, {Macri}, {Scolnic}, {Brout},
  {Casertano}, {Jones}, {Murakami}, {Anand}, {Breuval}, {Brink}, {Filippenko},
  {Hoffmann}, {Jha}, {D'arcy Kenworthy}, {Mackenty}, {Stahl}, \&
  {Zheng}}]{Riess_2022_hubblec}
{Riess}, A.~G., {Yuan}, W., {Macri}, L.~M., {et~al.} 2022, \apjl, 934, L7

\bibitem[{{Rodriguez-Gomez} {et~al.}(2019){Rodriguez-Gomez}, {Snyder}, {Lotz},
  {Nelson}, {Pillepich}, {Springel}, {Genel}, {Weinberger}, {Tacchella},
  {Pakmor}, {Torrey}, {Marinacci}, {Vogelsberger}, {Hernquist}, \&
  {Thilker}}]{TNG_sizes_Rodriguez_2019}
{Rodriguez-Gomez}, V., {Snyder}, G.~F., {Lotz}, J.~M., {et~al.} 2019, \mnras,
  483, 4140

\bibitem[{{Rojas} {et~al.}(2022){Rojas}, {Savary}, {Cl{\'e}ment}, {Maus},
  {Courbin}, {Lemon}, {Chan}, {Vernardos}, {Joseph}, {Ca{\~n}ameras}, \&
  {Galan}}]{Rojas_2022}
{Rojas}, K., {Savary}, E., {Cl{\'e}ment}, B., {et~al.} 2022, \aap, 668, A73

\bibitem[{{Romanowsky} \& {Kochanek}(1999)}]{Romanowsky_1999}
{Romanowsky}, A.~J. \& {Kochanek}, C.~S. 1999, \apj, 516, 18

\bibitem[{{Schneider} \& {Sluse}(2013)}]{Schneider2013}
{Schneider}, P. \& {Sluse}, D. 2013, \aap, 559, A37

\bibitem[{{Schneider} \& {Sluse}(2014)}]{Schneider2014}
{Schneider}, P. \& {Sluse}, D. 2014, \aap, 564, A103

\bibitem[{{Serjeant}(2014)}]{Serjeant14}
{Serjeant}, S. 2014, \apjl, 793, L10

\bibitem[{{Shajib} {et~al.}(2023){Shajib}, {Mozumdar}, {Chen}, {Treu},
  {Cappellari}, {Knabel}, {Suyu}, {Bennert}, {Frieman}, {Sluse}, {Birrer},
  {Courbin}, {Fassnacht}, {Villafa{\~n}a}, \& {Williams}}]{Shajib2023}
{Shajib}, A.~J., {Mozumdar}, P., {Chen}, G. C.~F., {et~al.} 2023, \aap, 673, A9

\bibitem[{{Sharma} \& {Linder}(2022)}]{Sharma_2022}
{Sharma}, D. \& {Linder}, E.~V. 2022, \jcap, 2022, 033

\bibitem[{{Shu} {et~al.}(2016){Shu}, {Bolton}, {Mao}, {Kochanek},
  {P{\'e}rez-Fournon}, {Oguri}, {Montero-Dorta}, {Cornachione},
  {Marques-Chaves}, {Zheng}, {Brownstein}, \& {M{\'e}nard}}]{Yiping2016}
{Shu}, Y., {Bolton}, A.~S., {Mao}, S., {et~al.} 2016, \apj, 833, 264

\bibitem[{{Shu} {et~al.}(2017){Shu}, {Brownstein}, {Bolton}, {Koopmans},
  {Treu}, {Montero-Dorta}, {Auger}, {Czoske}, {Gavazzi}, {Marshall}, \&
  {Moustakas}}]{Yiping_2017}
{Shu}, Y., {Brownstein}, J.~R., {Bolton}, A.~S., {et~al.} 2017, \apj, 851, 48

\bibitem[{{Shu} {et~al.}(2022){Shu}, {Ca{\~n}ameras}, {Schuldt}, {Suyu},
  {Taubenberger}, {Inoue}, \& {Jaelani}}]{Yiping_2022}
{Shu}, Y., {Ca{\~n}ameras}, R., {Schuldt}, S., {et~al.} 2022, \aap, 662, A4

\bibitem[{{Sonnenfeld} {et~al.}(2013{\natexlab{a}}){Sonnenfeld}, {Gavazzi},
  {Suyu}, {Treu}, \& {Marshall}}]{Sonnenfeld_2013_SL2S_photo}
{Sonnenfeld}, A., {Gavazzi}, R., {Suyu}, S.~H., {Treu}, T., \& {Marshall},
  P.~J. 2013{\natexlab{a}}, \apj, 777, 97

\bibitem[{{Sonnenfeld} {et~al.}(2013{\natexlab{b}}){Sonnenfeld}, {Treu},
  {Gavazzi}, {Suyu}, {Marshall}, {Auger}, \& {Nipoti}}]{Sonnenfeld_SL2S_2013}
{Sonnenfeld}, A., {Treu}, T., {Gavazzi}, R., {et~al.} 2013{\natexlab{b}}, \apj,
  777, 98

\bibitem[{{Sonnenfeld} {et~al.}(2015){Sonnenfeld}, {Treu}, {Marshall}, {Suyu},
  {Gavazzi}, {Auger}, \& {Nipoti}}]{Sonnenfeld_SL2S_2015_0884}
{Sonnenfeld}, A., {Treu}, T., {Marshall}, P.~J., {et~al.} 2015, \apj, 800, 94

\bibitem[{{Spiniello} {et~al.}(2015){Spiniello}, {Barnab{\`e}}, {Koopmans}, \&
  {Trager}}]{Spiniello_IMF_2015}
{Spiniello}, C., {Barnab{\`e}}, M., {Koopmans}, L.~V.~E., \& {Trager}, S.~C.
  2015, \mnras, 452, L21

\bibitem[{{Springel} {et~al.}(2018){Springel}, {Pakmor}, {Pillepich},
  {Weinberger}, {Nelson}, {Hernquist}, {Vogelsberger}, {Genel}, {Torrey},
  {Marinacci}, \& {Naiman}}]{IllustrisTNG_clustering_2018}
{Springel}, V., {Pakmor}, R., {Pillepich}, A., {et~al.} 2018, \mnras, 475, 676

\bibitem[{{Springel} {et~al.}(2001){Springel}, {Yoshida}, \&
  {White}}]{Springel_subfind_01}
{Springel}, V., {Yoshida}, N., \& {White}, S. D.~M. 2001, \na, 6, 79

\bibitem[{{Suyu} {et~al.}(2013){Suyu}, {Auger}, {Hilbert}, {Marshall}, {Tewes},
  {Treu}, {Fassnacht}, {Koopmans}, {Sluse}, {Blandford}, {Courbin}, \&
  {Meylan}}]{Suyu_2013}
{Suyu}, S.~H., {Auger}, M.~W., {Hilbert}, S., {et~al.} 2013, \apj, 766, 70

\bibitem[{{Suyu} {et~al.}(2010){Suyu}, {Marshall}, {Auger}, {Hilbert},
  {Blandford}, {Koopmans}, {Fassnacht}, \&
  {Treu}}]{Sherry_rectangular_vdisp_2010}
{Suyu}, S.~H., {Marshall}, P.~J., {Auger}, M.~W., {et~al.} 2010, \apj, 711, 201

\bibitem[{{Suyu} {et~al.}(2006){Suyu}, {Marshall}, {Hobson}, \&
  {Blandford}}]{Suyu_2006}
{Suyu}, S.~H., {Marshall}, P.~J., {Hobson}, M.~P., \& {Blandford}, R.~D. 2006,
  \mnras, 371, 983

\bibitem[{{Tessore} {et~al.}(2016){Tessore}, {Bellagamba}, \&
  {Metcalf}}]{Tessore2016}
{Tessore}, N., {Bellagamba}, F., \& {Metcalf}, R.~B. 2016, \mnras, 463, 3115

\bibitem[{{Thomas} {et~al.}(2013){Thomas}, {Steele}, {Maraston}, {Johansson},
  {Beifiori}, {Pforr}, {Str{\"o}mb{\"a}ck}, {Tremonti}, {Wake}, {Bizyaev},
  {Bolton}, {Brewington}, {Brownstein}, {Comparat}, {Kneib}, {Malanushenko},
  {Malanushenko}, {Oravetz}, {Pan}, {Parejko}, {Schneider}, {Shelden},
  {Simmons}, {Snedden}, {Tanaka}, {Weaver}, \& {Yan}}]{Portsmouth_Stellar_mass}
{Thomas}, D., {Steele}, O., {Maraston}, C., {et~al.} 2013, \mnras, 431, 1383

\bibitem[{{Tran} {et~al.}(2022){Tran}, {Harshan}, {Glazebrook}, {Keerthi
  Vasan}, {Jones}, {Jacobs}, {Kacprzak}, {Barone}, {Collett}, {Gupta},
  {Henderson}, {Kewley}, {Lopez}, {Nanayakkara}, {Sanders}, \&
  {Sweet}}]{Tran_2022}
{Tran}, K.-V.~H., {Harshan}, A., {Glazebrook}, K., {et~al.} 2022, \aj, 164, 148

\bibitem[{{Treu} \& {Marshall}(2016)}]{Treu_2016}
{Treu}, T. \& {Marshall}, P.~J. 2016, \aapr, 24, 11

\bibitem[{{Vagnozzi} {et~al.}(2022){Vagnozzi}, {Pacucci}, \&
  {Loeb}}]{Vagnozzi_2022_Hubblec}
{Vagnozzi}, S., {Pacucci}, F., \& {Loeb}, A. 2022, Journal of High Energy
  Astrophysics, 36, 27

\bibitem[{{van de Ven} {et~al.}(2010){van de Ven}, {Falc{\'o}n-Barroso},
  {McDermid}, {Cappellari}, {Miller}, \& {de Zeeuw}}]{van_de_Ven_2010}
{van de Ven}, G., {Falc{\'o}n-Barroso}, J., {McDermid}, R.~M., {et~al.} 2010,
  \apj, 719, 1481

\bibitem[{{Van de Vyvere} {et~al.}(2022){Van de Vyvere}, {Gomer}, {Sluse},
  {Xu}, {Birrer}, {Galan}, \& {Vernardos}}]{Lyne_2022}
{Van de Vyvere}, L., {Gomer}, M.~R., {Sluse}, D., {et~al.} 2022, \aap, 659,
  A127

\bibitem[{{Vegetti} {et~al.}(2010){Vegetti}, {Koopmans}, {Bolton}, {Treu}, \&
  {Gavazzi}}]{Vegetti_2010}
{Vegetti}, S., {Koopmans}, L.~V.~E., {Bolton}, A., {Treu}, T., \& {Gavazzi}, R.
  2010, \mnras, 408, 1969

\bibitem[{{Velliscig} {et~al.}(2014){Velliscig}, {van Daalen}, {Schaye},
  {McCarthy}, {Cacciato}, {Le Brun}, \& {Dalla Vecchia}}]{Velliscig_2014}
{Velliscig}, M., {van Daalen}, M.~P., {Schaye}, J., {et~al.} 2014, \mnras, 442,
  2641

\bibitem[{{Wang} {et~al.}(2019){Wang}, {Vogelsberger}, {Xu}, {Shen}, {Mao},
  {Barnes}, {Li}, {Marinacci}, {Torrey}, {Springel}, \&
  {Hernquist}}]{Wang_2019_ETG_profiles_II}
{Wang}, Y., {Vogelsberger}, M., {Xu}, D., {et~al.} 2019, \mnras, 490, 5722

\bibitem[{{Warren} \& {Dye}(2003)}]{Warren_2003}
{Warren}, S.~J. \& {Dye}, S. 2003, \apj, 590, 673

\bibitem[{{White} \& {Frenk}(1991)}]{White_91_ETG}
{White}, S. D.~M. \& {Frenk}, C.~S. 1991, \apj, 379, 52

\bibitem[{{Wong} {et~al.}(2020){Wong}, {Suyu}, {Chen}, {Rusu}, {Millon},
  {Sluse}, {Bonvin}, {Fassnacht}, {Taubenberger}, {Auger}, {Birrer}, {Chan},
  {Courbin}, {Hilbert}, {Tihhonova}, {Treu}, {Agnello}, {Ding}, {Jee},
  {Komatsu}, {Shajib}, {Sonnenfeld}, {Blandford}, {Koopmans}, {Marshall}, \&
  {Meylan}}]{Wong_2020}
{Wong}, K.~C., {Suyu}, S.~H., {Chen}, G. C.~F., {et~al.} 2020, \mnras, 498,
  1420

\bibitem[{{Y{\i}ld{\i}r{\i}m} {et~al.}(2020){Y{\i}ld{\i}r{\i}m}, {Suyu}, \&
  {Halkola}}]{Akin_2020}
{Y{\i}ld{\i}r{\i}m}, A., {Suyu}, S.~H., \& {Halkola}, A. 2020, \mnras, 493,
  4783

\end{thebibliography}

%\printbibliography

\end{document}